\documentclass[a4paper,11pt]{article}
\usepackage[utf8]{inputenc}
\usepackage[english]{babel}
\usepackage{braket}
\usepackage{dsfont}
\usepackage{bbm}
\usepackage{mathdots}
\usepackage{stackrel}
\usepackage{xcolor}
\usepackage{mathtools}
\usepackage{bbm}
\usepackage{subfigure}
\usepackage[T1]{fontenc}               % codifica dei font
\usepackage[left=3cm,
            right=2.5cm,
            top=2.5cm,
            bottom=3cm
            ]{geometry}                % spaziatura bordi
\usepackage{graphicx}
\usepackage{appendix}
\usepackage{fancyhdr}
\usepackage{amsmath,                   % simboli matematici
            amssymb,                   % altri simboli matematici
            amsthm}                    % stili teorema
\usepackage[margin=30pt, bf, font=small, center, justification=justified]{caption}[2004/07/16]

\newif\ifnatbibsort\natbibsorttrue

\relax

\RequirePackage[colorlinks=true,urlcolor=blue,anchorcolor=blue,citecolor=blue,filecolor=blue,linkcolor=blue,menucolor=blue,pagecolor=blue,linktocpage=true,pdfproducer=medialab,pdfa=true]{hyperref}

\newcommand{\gcal}
{\mathcal{G}}

\title{\bf Entangled multiplets, asymmetry, and quantum Mpemba effect in dissipative systems}

\author{Fabio Caceffo$^{1}$, Sara Murciano$^{2,3}$ and Vincenzo Alba$^{1}$}

\date{}

\begin{document} 

\maketitle
{\small
\vspace{-5mm}  \ \\
{$^{1}$}  Dipartimento di Fisica dell'Universit\`a di Pisa and INFN, Sezione di Pisa, I-56127 Pisa, Italy\\[0.1cm]
\medskip
$^{2}$	Walter Burke Institute for Theoretical Physics, Caltech, Pasadena, CA 91125, USA\\
\medskip
$^{3}$	Department of Physics and IQIM, Caltech, Pasadena, CA 91125, USA\\[-0.1cm]
\medskip
}

\begin{abstract}

Recently, the entanglement asymmetry emerged as an informative tool to understand 
dynamical symmetry restoration in out-of-equilibrium quantum many-body systems after a quantum 
quench. 
For integrable systems the asymmetry can be understood in the 
space-time scaling limit via the quasiparticle picture, as it was pointed out in Ref.~\cite{ares2023entanglement}. 
However, a quasiparticle picture for quantum quenches from generic initial states was still lacking. 
Here we conjecture a full-fledged quasiparticle picture for the charged moments of the reduced 
density matrix, which are the main ingredients to construct the asymmetry. Our formula works for  
quenches producing entangled multiplets of an arbitrary number of excitations. We benchmark our results 
in the $XX$ spin chain. First, by using an elementary approach based on the multidimensional stationary phase 
approximation we provide an \emph{ab initio} rigorous derivation of the dynamics of the charged moments for the quench treated in~\cite{ares2023lack}. Then, 
we show that the same results can be straightforwardly obtained within our quasiparticle picture. 
As a byproduct of our analysis, we obtain a general criterion ensuring a vanishing entanglement asymmetry at long times.
Next, by using the Lindblad master equation, we study the effect of gain and loss dissipation on the 
entanglement asymmetry. Specifically, we investigate the fate of the so-called quantum Mpemba effect (QME) in the 
presence of dissipation. We show that dissipation can induce QME even if unitary 
dynamics does not show it, and we provide a quasiparticle-based interpretation of the condition for the QME.

\end{abstract}

 %\tableofcontents

 \newpage

\section{Introduction}

\begin{figure}[t!]
\includegraphics[width=0.95\textwidth]{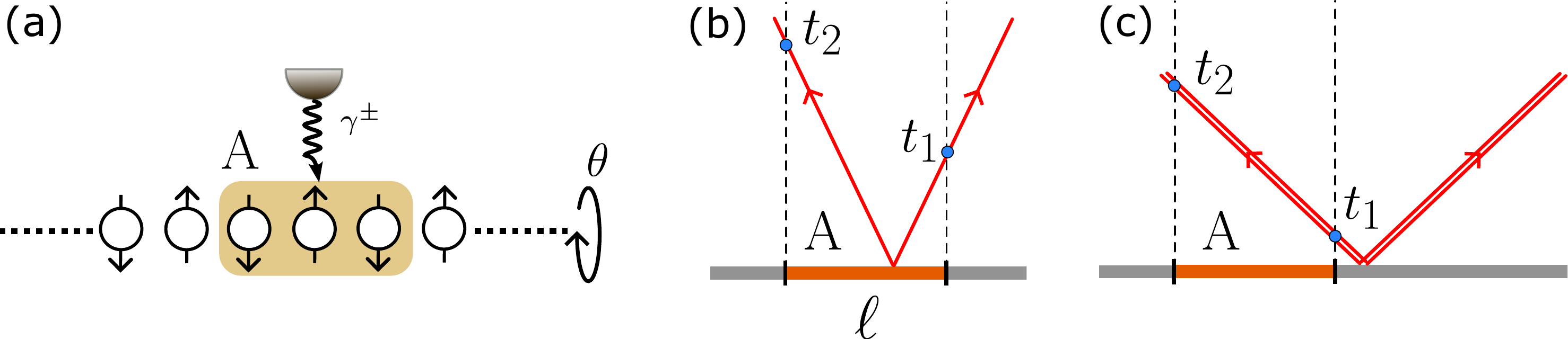}
     \caption{Setup considered in this work. (a) Quench from the tilted N\'eel state, with $\theta$ the 
     tilting angle, in the $XX$ chain in the presence of global gain and loss dissipation. We are interested in 
     the dynamics of the entanglement asymmetry of a finite region $A$ embedded in an infinite chain. (b)  Quasiparticle picture 
     for the asymmetry: the case of entangled pairs. If entangled pairs of quasiparticles are produced after the quench, only the 
     pairs that are fully contained in the subsystem contribute. Here the pair contributes for $t\le t_1$.  At longer times the pair is not fully contained in $A$, and the asymmetry vanishes. (c) For the tilted N\'eel state quadruplets of entangled quasiparticles are produced with velocities $v_1=v_2$ and $v_3=v_4$.  In contrast with (b), in the absence of dissipation the asymmetry does not 
     vanish at long times because there are always paired fermions in $A$. 
     }
     \label{fig:cartoon}
\end{figure}

Symmetries are often used by physicists as Lego blocks upon which they build our understanding of Nature and its fundamental laws. 
Indeed, the concepts of symmetry and symmetry-breaking are ubiquitous in Physics. In recent years there has been renewed interest in 
studying symmetry-breaking both in and out-of-equilibrium systems~\cite{ares2023entanglement,ares2023lack,capizzi2023entanglement,capizzi2023universal,rylands2023microscopic,ferro2023nonequilibrium,chen2023entanglement,ares2023page,murciano2023entanglement,khor2023confinement,joshi2024observing}. Let us consider the prototypical setup of the quantum quench~\cite{calabrese2016introduction} in which a system, here a one-dimensional one, is prepared in an initial state and 
let to evolve under a many-body Hamiltonian $H$. Let us also assume that the Hamiltonian commutes with a charge operator $Q$, which 
generates an Abelian symmetry $U(1)$. One can consider the situation in which the initial state breaks explicitly the symmetry. An intriguing consequence of thermalization, or equilibration to a  Generalized Gibbs Ensemble~\cite{calabrese2016introduction} ($GGE$)  
is that the symmetry is, at least for typical initial states (see Ref.~\cite{fagotti2014on}), dynamically restored at long times~\cite{fagotti2014relaxation}. Very recently, the question how fast a symmetry is restored as a function of the degree of symmetry breaking in the initial state, and how this is reflected in entanglement-related quantities attracted a lot of attention. The so-called \emph{entanglement asymmetry}~\cite{ares2023entanglement} 
emerged as an informative tool. The asymmetry (see section~\ref{sec:defs}) is defined from the reduced density matrix $\rho_A$ of a subsystem $A$ (see Fig.~\ref{fig:cartoon}). Let us consider the case of a $U(1)$ charge operator $Q=Q_A+Q_{\bar A}$, with $A$ a finite region, and 
$\bar A$ its complement. Since the initial state breaks the $U(1)$ symmetry of the model, $\rho_A$ does not commute with the 
charge $Q_A$ restricted to the subsystem, implying that $\rho_A$ does not exhibit a block-diagonal structure. This means that if one insisted on using a basis of eigenstates of $Q_A$, $\rho_A$ would have matrix elements between states with different 
eigenvalues of $Q_A$. The asymmetry is sensitive to these matrix elements. If the symmetry is dynamically restored,  
matrix element connecting different charge sectors, and hence the entanglement asymmetry, vanish at long times. 
Surprisingly, it was observed that in several quenches the more the symmetry is broken in the initial state, the faster is restored~\cite{ares2023entanglement}. 
This provides a counterpart of the classical \emph{Mpemba effect} in out-of-equilibrium quantum many-body systems.  
The classical version of the Mpemba effect can be summarised in the sentence “hot water freezes faster than cold water”, since it was primarily observed in water~\cite{mpemba1969cool}. 
Classical Mpemba effect was observed also in several systems, such as  clathrate 
hydrates~\cite{ahn2016experimental}, magnetic alloys~\cite{chaddah2010overtaking}, carbon nanotube resonators~\cite{greaney2011mpemba}, 
granular gases~\cite{lasanta2017when}, colloidal systems~\cite{kumar2020exponentially} or dilute atomic gases \cite{keller2018quenches}. 
A microscopic explanation of the quantum Mpemba effect in both interacting and free integrable spin chains has been provided in~\cite{rylands2023microscopic} (see also~\cite{murciano2023entanglement} for a further study of the asymmetry in free integrable quantum systems).

Very recently, the asymmetry was measured experimentally by using a trapped-ion quantum simulator, and the quantum Mpemba effect was confirmed~\cite{joshi2024observing} (see also 
Ref.~\cite{Chatterjee2023,shapira2024mpemba,zhang2024observation,wang2024mpemba} for the observation of another version of the quantum Mpemba effect in a  single qubit). 
The dynamics of the entanglement asymmetry has been studied for a quench from the tilted ferromagnetic state~\cite{ares2023entanglement}, and, more in general, from the ground state of the $XY$ spin chain~\cite{murciano2023entanglement}, which does not preserve the transverse magnetization. In these situations, the dynamics determined by a free $U(1)$-invariant Hamiltonian leads to the symmetry restoration of the $U(1)$ charge in the subsystem. 
Crucially, the dynamics of the entanglement asymmetry, and the onset of the quantum Mpemba effect,  can be understood 
in the space-time scaling limit (hydrodynamic limit) $\ell,t\to\infty$ with the ratio $t/\ell$ fixed. Indeed, 
the main formula obtained in Refs.~\cite{ares2023entanglement,murciano2023entanglement} is reminiscent of the so-called quasiparticle picture~\cite{calabrese2005evolution,fagotti2008evolution,alba2017entanglement,calabrese2018entanglement} for 
the entanglement entropy. Still, a full-fledged quasiparticle picture was not available so far. For instance, a general prescription to determine  how different configurations of entangled quasiparticles  contribute to the asymmetry was not completely understood.

Here we first review the results of Refs.~\cite{ares2023entanglement,ares2023lack,murciano2023entanglement}. We focus on a dynamics governed by the $XX$ chain, considering quantum quenches from the ground state of the $XY$ chain and the tilted N\'eel state.  In this setup, the $U(1)$ charge corresponds to the total magnetization, which commutes with the Hamiltonian of the $XX$ chain.  On the other hand, 
both the initial states that we consider do not have a well-defined total magnetization, and hence break the $U(1)$ symmetry. The dynamics after the quench 
from the ground state of the $XY$ chain is understood in terms of the ballistic propagation of pairs of entangled quasiparticles, 
whereas the quench from the tilted N\'eel state gives rise to entangled \emph{quadruplets} of quasiparticles. Interestingly, while in the former case the symmetry is restored at long times, and the entanglement asymmetry vanishes, in the latter this does not happen. This is due to the fact that the $XX$ chain possesses some non-Abelian conservation laws~\cite{ares2023lack,fagotti2014on}. 
Specifically, we show that the framework of entangled 
multiplets developed in Refs.~\cite{bertini2018entanglementevolution,bastianello2018spreading,caceffo2023negative} is the natural 
one  to build the quasiparticle picture for the entanglement asymmetry. First, by using multidimensional stationary phase methods 
we provide an \emph{ab initio} derivation of the results of Ref.~\cite{ares2023lack}. In contrast with Ref.~\cite{ares2023entanglement} (see also Ref.~\cite{ares2023lack}) our derivation does not rely heavily on the 
Toeplitz structure of the two-point fermionic correlation function. Based on these results, we conjecture the quasiparticle picture 
for the charged moments of the reduced density matrix, which are the main ingredients to obtain the entanglement 
asymmetry.  Our formula is  by construction  readily generalized to quantum quenches generating entangled multiplets of arbitrary size. 
As in the standard quasiparticle picture, the dynamics of the charged moments and the asymmetry depend  on the way the 
members of the entangled multiplet are shared between the subsystem $A$ and the rest. 
Crucially, in our generalized quasiparticle picture the entanglement asymmetry receives nonzero contributions only from 
configurations in which at least two entangled quasiparticles are in the subsystem (see Fig.~\ref{fig:cartoon}) (b). This suggests that in generic quenches governed by the XX Hamiltonian producing multiplets of arbitrary size, the symmetry is always restored at long times. This happens because at long times any two quasiparticles in the multiplet cannot be in $A$ at the same time, provided that the quasiparticles forming the multiplet have different velocities. Interestingly, the last condition is violated in the quench from the tilted N\'eel state, which gives rise to quadruplets of quasiparticles with velocities $v_1(k)=v_2(k)$ and $v_3(k)=v_4(k)$, with $k$ the quasimomentum.  This means that at long time there is always a pair of entangled quasiparticles in $A$ (see Fig.~\ref{fig:cartoon} (c)). This prevents the symmetry restoration, which is reflected in a nonzero entanglement asymmetry. 

Next we study the effect of dissipation on the quantum Mpemba effect. Indeed, the question as to whether the Mpemba effect 
persists in the presence of an environment has attracted recent attention. For instance, in Ref.~\cite{joshi2024observing} 
it was shown that Mpemba effect does not occur under pure dephasing dynamics. On the other hand, it has been shown that 
Mpemba effect can occur in Markovian dynamics~\cite{nava2019lindblad,carollo2021exponentially,shapira2024mpemba}. 
We focus on the dynamics starting from the tilted N\'eel state in the $XX$ chain in the presence of gain and loss dissipation, 
which describe incoherent emission or absorption of fermions from the chain. We employ the framework of the Lindblad 
master equation~\cite{petruccione2002the}. Crucially, for the gain and loss dissipation the Lindblad operators are linear, so that the Gaussianity of the initial state is preserved and we can study the quantum Mpemba effect analytically. 
In contrast with the unitary dynamics, which does not support symmetry restoration and Mpemba effect, we show that 
in the presence of gain/loss dissipation both symmetry restoration and quantum Mpemba effect can occur. We also derive 
the conditions that give rise to the Mpemba effect, comparing them with the case without dissipation.

The manuscript is organised as follows. In section~\ref{sec:models} we introduce the $XY$ and
$XX$ chains. We also introduce the framework of the Lindblad equation and gain/loss dissipation. In section~\ref{sec:defs}, we introduce the entanglement asymmetry, and the charged moments of the reduced density matrix. In section~\ref{sec:leading}, 
we provide an \emph{ab initio} derivation of the charged moments in the hydrodynamic limit after the quench from the tilted N\'eel state 
in the $XX$ chain. In Section~\ref{subsec:nodiss} we consider the unitary dynamics, whereas in Section~\ref{subsec:diss} we 
discuss the effect of gain and loss dissipation. In Section~\ref{sec:chmom_quasipart} we conjecture a  generalized quasiparticle 
picture for the charged moments. In particular, in Section~\ref{sec:quasi-ent} we introduce the framework of entangled multiplets and the quasiparticle picture for 
the von Neumann entropy, discussing the generalization to the charged moments in Section~\ref{sec:mult-quasi}. In Section~\ref{sec:quench-XX},  by using the generalized quasiparticle picture, we rederive the formula describing the dynamics of the charged moments after the quench from the tilted N\'eel state in the $XX$  chain. In 
Section~\ref{sec:quench-XY} we obtain the charged moments after the quench from the ground state of the 
$XY$  chain that was derived in Ref.~\cite{murciano2023entanglement}.  In Section~\ref{sec:results} we discuss the effect of gain and loss 
dissipation on the quantum Mpemba effect, focusing on the quench from the tilted N\'eel state in the $XX$ chain. In particular, 
we discuss the condition ensuring the onset of Mpemba effect. In Section~\ref{sec:quasi-mpemba} we compare that condition 
with the case without dissipation. In Section~\ref{sec:numerics} we discuss numerical benchmarks for the results of 
Section~\ref{sec:results}. Specifically, 
in Section~\ref{sec:charged-num} we discuss the dynamics of the charged moments, whereas in Section~\ref{sec:asy-num} we 
consider the entanglement asymmetry. We conclude in Section~\ref{sec:concl}. Finally, in Appendix~\ref{app:proof_nodiss} 
we detail the derivation of some of the results of Section~\ref{sec:leading}. Some generalizations to the dissipative 
case are discussed in Appendix~\ref{app:diss}. In Appendix~\ref{app:XY_proof} we provide the proof of some of the 
formulas of Section~\ref{sec:chmom_quasipart}. 

\section{Quantum quenches in the $XY$ chain  with  dissipation}\label{sec:models}

Here we deal with  the out-of-equilibrium dynamics after a quantum quench in one-dimensional 
integrable systems~\cite{calabrese2016introduction}. Given an initial state $|\Psi(0)\rangle$ and a Hamiltonian $H$, the 
time-evolved state $\Psi(t)$ is given  as 
\begin{equation}\label{eq:quench}
 \ket{\Psi(t)}= e^{-i tH}\ket{\Psi(0)}.
\end{equation}
We focus on the  XY spin chain given in terms of the Hamiltonian $H$ as
\begin{equation}
    \label{eq:XYchain-spin}
    H_{XY}=-\frac{1}{4}\sum_{j=-\infty}^\infty \left((1+\gamma)\sigma^x_j \sigma^x_{j+1} + (1-\gamma) \sigma^y_j  \sigma^y_{j+1}  - 2h \sigma^z_j\right),
\end{equation}
where $\sigma^{x,y,z}_j$ are Pauli matrices acting on site $j$ of the chain, and $h$ is a magnetic field. 
The chain is infinite. After a Jordan-Wigner transformation, Eq.~\eqref{eq:XYchain-spin} is mapped to a 
quadratic fermionic Hamiltonian as 
\begin{equation}
    \label{eq:XYchain}
    H_{XY}=-\frac{1}{2}\sum_{j=-\infty}^\infty \left(c_j^\dagger c_{j+1} + \gamma c_j^\dagger  c_{j+1}^\dagger + \text{h.c.} - 2h c_j^\dagger c_j\right),
\end{equation}
where $c_j,c_j^\dagger$ are standard fermionic operators. 
Eq.~\eqref{eq:XYchain} is diagonalized by first rewriting it in terms of the  Fourier transformed fermionic operators  
$c_k:=\sum_{j=-\infty}^\infty e^{-ikj}c_j$, and then by applying a Bogoliubov transformation as 
\begin{equation}
    \label{eq:Bogoliubov}
    \begin{pmatrix}
        \eta_k\\\eta_{2\pi-k}^\dagger
    \end{pmatrix}=
    \begin{pmatrix}
        \cos(\Delta_k/2)&-i \sin(\Delta_k/2)\\ -i\sin(\Delta_k/2)&\cos(\Delta_k/2)
    \end{pmatrix}
    \begin{pmatrix}
        c_k\\c_{2\pi-k}^\dagger
    \end{pmatrix}.
\end{equation}
Here, $\eta_k,\eta^\dagger_k$  are new fermionic operators, and 
the Bogoliubov angle $\Delta_k$ is defined as 
\begin{equation}
    \label{eq:Bogoliubov_angle}
    \cos \Delta_k.:=\frac{h-\cos k}{\sqrt{(h-\cos k)^2+\gamma^2 \sin^2 k}}, \qquad \sin \Delta_k:=\frac{\gamma \sin k}{\sqrt{(h-\cos k)^2+\gamma^2 \sin^2 k}}. 
\end{equation}
The Hamiltonian~\eqref{eq:XYchain} in terms of $\eta_k$ is diagonal as  
\begin{equation}
    \label{eq:XY_diag}
    H_{XY}=\int_0^\pi \frac{dk}{2\pi} \varepsilon_{XY}(k) (\eta_k^\dagger \eta_k-\eta_{2\pi-k}\eta_{2\pi-k}^\dagger),
\end{equation}
with $\varepsilon_{XY}(k):=\sqrt{(h-\cos k)^2+\gamma^2 \sin^2 k}$ the single-particle dispersion. 
The ground state (not normalized) of the XY spin chain is
\begin{equation}
    \label{eq:XY_ground}
    |\Omega\rangle = \prod_k \eta_k |0\rangle,
\end{equation}
where $|0\rangle$ is the state annihilated by the Dirac operators $c_j$, and satisfies $\eta_k|\Omega\rangle=0 \; \forall k$.

For $\gamma=0$, the Hamiltonian~\eqref{eq:XYchain-spin} becomes that of the $XX$ spin chain given as 
\begin{equation}\label{eq:xx}
 H=-\frac{1}{4}\sum_{j=-\infty}^{\infty}\left[\sigma_{j}^x\sigma_{j+1}^x
 +\sigma_{j}^y\sigma_{j+1}^y-2h\sigma_j^z\right],
\end{equation}
The fermionic Hamiltonian~\eqref{eq:XYchain} for $\gamma=h=0$ becomes the tight-binding free-fermion chain 
\begin{equation}
\label{eq:xx-free}
H_{XX}=-\frac{1}{2}\sum_{j=-\infty}^\infty\left[c^\dagger_jc_{j+1}+c^\dagger_{j+1}c_j\right]. 
\end{equation}
Eq.~\eqref{eq:xx-free} can be  diagonalised in terms of  the Fourier transformed operators  
$c_k$ (cf.~\eqref{eq:Bogoliubov}) as
\begin{equation}
    \label{eq:XX_diag}
    H_{XX}=\int_0^{2\pi}\frac{dk}{2\pi} \varepsilon_{XX}(k) c_k^\dagger c_k,
\end{equation}
where the one-particle dispersion relation is $\varepsilon_{XX}(k)=-\cos(k)$. 
For the following, it is important to stress that 
$H_{XX}$, unlike the $XY$ chain~\eqref{eq:XYchain-spin}, preserves the transverse global magnetization, i.e., the~\eqref{eq:xx} 
satisfies $[H_{XX}, Q]=0$, with  
\begin{equation}\label{eq:trans_mag}
 Q=\frac{1}{2}\sum_{j}\sigma_j^z.
\end{equation}
In the fermionic language Eq.~\eqref{eq:trans_mag} can be written as 
$Q=\sum_{j} (c_j^\dagger c_j -1/2)$. 
For the following, it is also crucial to observe that the Hamiltonian~\eqref{eq:XY_diag} of the $XY$ chain  
can be written in terms of $m$ different species of quasiparticles, each having a dispersion relation 
$\varepsilon_j(k)$ ($j=1,\dots, m$), as
\begin{equation}
        \label{eq:gen-quasipart_1}
        H=\int_\mathcal{B}\frac{dk}{2\pi} \sum_{j=1}^m \varepsilon_j(k) d_j^\dagger(k)d_j(k).
\end{equation}
Here $d_j(k)$ are the fermionic operators associated to the different species of quasiparticles and satisfy the usual anticommutation relations $\{d_i(k),d^{\dagger}_j(p)\}=\delta_{ij}\delta(k-p)$, while $\mathcal{B}$ is the reduced Brillouin zone $\mathcal{B}\subseteq [0,2\pi]$. The operators $d_j(k)$ are obtained from 
the operators $\eta(k)$ diagonalizing~\eqref{eq:XYchain-spin}. 
 The number of species is determined by the prequench initial state. The reason for the rewriting~\eqref{eq:gen-quasipart_1} is because the prequench initial state $|0\rangle$ can induce 
non-diagonal correlations in quasimomentum space, i.e., $\langle0|\eta^\dagger(k)\eta(k')|0\rangle\ne \delta_{k,k'}$. This means that  different species of quasiparticles  share nontrivial correlations. Indeed, 
typically the quasiparticles of different species are entangled, forming an entangled multiplet. 
These entangled multiplets are essential to accurately describe the dynamics of entanglement-related quantities~\cite{bastianello2018spreading,caceffo2023negative,bertini2018entanglement}.

Here we consider the dynamics starting from the initial Gaussian state $\Psi(0)$, defined as  
\begin{equation}\label{eq:cat}
\ket{\Psi(0)}:=\frac{\ket{ \theta}-\ket{-\theta}}{\sqrt{2}},
\end{equation}
where $\ket{\theta}$ is the tilted Néel state 
\begin{equation}\label{eq:tiltedNéel}
\ket{ \theta}:=e^{i\frac{\theta}{2} \sum_{j}\sigma_j^y}|\downarrow\uparrow\cdots\rangle.
\end{equation}
Crucially, for $\theta\neq m \pi, m \in \mathbb{Z}$, it breaks the $U(1)$ symmetry associated with 
the conservation of the transverse magnetisation~\eqref{eq:trans_mag}. 

A crucial quantity to determine entanglement properties~\cite{peschel2009reduced} of the $XY$ chain is the 
two-point correlation matrix $\Gamma_{jj'}(t)$ defined as 
\begin{equation}\label{eq:two_point_corr}
\Gamma_{jj'}(t):=2\bra{\Psi(t)}
    \boldsymbol{c}_j^\dagger
    \boldsymbol{c}_{j'}
    \ket{\Psi(t)}-\delta_{j,j'},\quad j, j'\in A, 
\end{equation}
where $\boldsymbol{c}_j:=(c_j,c_j^\dagger)$,  and 
$c_j$ are the fermionic operators appearing in~\eqref{eq:xx-free}.

In this work we are interested in understanding the effect of dissipation on the 
quantum Mpemba effect. To account for dissipation, we employ the framework of the 
Lindblad master equation~\cite{petruccione2002the}.  
The Lindblad equation describes the dynamics of the full-system density matrix $\rho(t)$ as  
\begin{equation}
\label{eq:lind}
\frac{d\rho_t}{dt}=\mathcal{L}(\rho(t)):=-i[H,\rho(t)]+\sum_{j=1}^L\sum_{\alpha=\pm}\left(L_{j,\alpha}\rho(t) 
L_{j,\alpha}^\dagger-\frac{1}{2}\left\{L_{j,\alpha}^\dagger L_{j,\alpha},\rho(t)\right\}\right)\, .
\end{equation}
Here $L_{j,\alpha}$ are the Lindblad jump operators encoding dissipation. 
We restrict ourselves to the paradigmatic case of gain and loss dissipation. 
They encode incoherent processes in which fermions are injected or removed in the 
chain at rates $\gamma^+$ and $\gamma^-$, respectively. 
The Lindblad jump operators, $L_{j,\alpha}$,
for gain and loss processes are defined as  
$L_{j,-}:=\sqrt{\gamma^-}c_j$ and $L_{j,+}:=\sqrt{\gamma^+}c_j^\dagger$. 
The generic Lindblad equation is not exactly solvable, not even for 
integrable Hamiltonians. Remarkably, if $H$ is quadratic in the fermion operators, and 
the Lindblad operators are linear in the fermions, Eq.~\eqref{eq:lind} can be 
solved analytically via the ``third quantization'' approach~\cite{prosen2008third}. 
Finally, it is crucial to observe that for gain and loss processes the Gaussianity of $\rho$ is preserved during the 
time evolution under Eq.~\eqref{eq:lind}.

\section{Entanglement asymmetry \& charged moments}\label{sec:defs}

Let us review the formal definition of the entanglement asymmetry~\cite{ares2023entanglement}, which is the main topic 
of this manuscript. We consider a pure state $\ket{\psi}$ living in the bipartition $A\cup \bar{A}$ (see Fig.~\ref{fig:cartoon} (a)). 
From here, one can build the reduced density matrix $\rho_A=\mathrm{Tr}_{\bar{A}}(\ket{\psi}\bra{\psi})$ describing the subsystem $A$. 
Let us consider a local charge operator $Q$, which generates a $U(1)$ symmetry for the 
full system, and it can be decomposed  as $Q=Q_A+Q_{\bar{A}}$, with $Q_A$ and $Q_{\bar{A}}$ the 
charge operator acting in subsystem $A$ and $\bar{A}$, respectively. If $\ket{\psi}$ is an eigenstate of $Q$, $[\rho_A,Q_A]=0$ and $\rho_A$ has a block-diagonal structure in the eigenbasis of $Q_A$, with each block labelled by an integer $q \in \mathbb{Z}$. If $[\rho_A,Q_A]\neq 0$, $\rho_A$ exhibits non-zero entries outside the blocks. Thus, we can define a new density 
matrix $\rho_{A,Q}:=\sum_q\Pi_q\rho_A\Pi_q$ by using the projectors on the eigenspace of $q$, $\Pi_q$, 
such that $[\rho_{A,Q},Q_A]=0$. 
Notice that if $[\rho_A,Q_A]=0$, then $\rho_{A,Q}=\rho_A$. On the other hand,
if $[\rho_A,Q_A]\ne0$, then $\rho_{A,Q}$ is obtained from $\rho_A$ by removing 
the matrix elements that connect blocks of $\rho_A$ with different values of $Q_A$. In analogy with the definitions of the R\'enyi entropies $S_A^{(n)}(\rho)=1/(1-n)\ln \mathrm{Tr}\rho^n$, we can introduce the R\'enyi entanglement asymmetry as 
\begin{equation}\label{eq:def_renyi}
    \Delta S_A^{(n)}:=S_A^{(n)}(\rho_{A,Q})-S_A^{(n)}(\rho_A),
\end{equation}
i.e., the difference between the R\'enyi entropy  obtained from $\rho_{A,Q}$ and the one obtained from  $\rho_{A}$. In the replica limit $n\to 1$, the R\'enyi entropies provide the von Neuman entropy, $S(\rho)=-\mathrm{Tr}(\rho \ln \rho)$, and $ \Delta S_A^{(n)}$ reduces to 
\begin{equation}
\label{eq:asy-first}
  \Delta S_A= S_A(\rho_{A,Q})-S_A(\rho_A).  
\end{equation}
The entanglement asymmetry satisfies two crucial properties. First,  $ \Delta S_A^{(n)}\geq 0$. Second, 
it vanishes if and only if the symmetry is not broken, i.e. $[\rho_A,Q_A]=0$. 

Before digging into the physics of the asymmetry, we introduce the charged moments,  which play a fundamental role to compute $\Delta S_A^{(n)}$. By exploiting the Fourier representation of the projector operator, $\Pi_q$, the reduced density matrix $\rho_{A,Q}$ reads
\begin{equation}
\label{eq:rhoa-1}
 \rho_{A, Q}=\int_{-\pi}^\pi \frac{d\alpha}{2\pi}e^{-i\alpha Q_A}\rho_A e^{i\alpha Q_A}. 
\end{equation}
From~\eqref{eq:rhoa-1} one obtains 
\begin{equation}\label{eq:FT}
 \mathrm{Tr}(\rho_{A, Q}^n)=\int_{-\pi}^\pi \frac{d\alpha_1\dots d\alpha_n}{(2\pi)^n} Z_n(\boldsymbol{\alpha}),
 \end{equation}
where $\boldsymbol{\alpha}=\{\alpha_1,\dots,\alpha_n\}$ and
\begin{equation}\label{eq:Znalpha}
 Z_n(\boldsymbol{\alpha}):=
 \mathrm{Tr}\left[\prod_{j=1}^n\rho_A e^{i\alpha_{j,j+1}Q_A}\right],
\end{equation}
with $\alpha_{ij}:=\alpha_i-\alpha_j$ and $\alpha_{n+1}\equiv\alpha_1$. 
The quantities $Z_n(\boldsymbol{\alpha})$ are known as charged moments and, 
by using~\eqref{eq:FT}, allow to obtain the entanglement asymmetry as
\begin{equation}
    \label{eq:asym_from_chmom}
    \Delta S_A^{(n)}=\frac{1}{1-n}\ln\left(\int_{-\pi}^\pi \frac{d\alpha_1\dots d\alpha_n}{(2\pi)^n} \frac{Z_n(\boldsymbol{\alpha})}{Z_n(\boldsymbol{0})}\right).
\end{equation}
Crucially, the asymmetry depends only on the ratio $\frac{Z_n(\boldsymbol{\alpha})}{Z_n(\boldsymbol{0})}$. 

Since at any time, both $\rho_A(t)$ and $Q_A=\sum_{j\in A} c^\dagger_j c_j$ are Gaussian operators in terms of the 
fermionic operators $\boldsymbol{c}_j=(c_j, c_j^\dagger)$, the charged moments can be obtained from the two-point correlation function $\Gamma(t)$ (cf.~\eqref{eq:two_point_corr}) restricted to the subsystem $A$~\cite{ares2023entanglement}. By using  the composition rules for the trace of a product of Gaussian operators~\cite{fagotti2014on}, the charged moments~\eqref{eq:Znalpha} can be written as~\cite{ares2023entanglement}
\begin{equation}
    \label{eq:chmom_gamma}   Z_n(\boldsymbol{\alpha},t)=\sqrt{\det\left[\left(\frac{\mathds{1}-\Gamma(t)}{2}\right)^n\left(\mathds{1}+\prod_{j=1}^n W_j(t)\right)\right]},
\end{equation}
where $W_j(t)=(\mathds{1}+\Gamma(t))/(\mathds{1}-\Gamma(t))e^{i(\alpha_j-\alpha_{j+1})N_A}$, with $N_A$ a diagonal matrix taking into account the presence of the charge operator $Q_A$ in~\eqref{eq:Znalpha} and such that $(N_A)_{2j-1,2j-1}=-1$ and $(N_A)_{2j,2j}=1$. 

\section{Hydrodynamic limit of the charged moments after generic quenches in the XX chain: A rigorous derivation}\label{sec:leading}

As we have explained in the previous section,  the charged moments in Eq.~\eqref{eq:Znalpha} are the starting point to compute the asymmetry. In the following we focus on the dynamics of the charged moments after a quantum quench. Specifically, we provide a rigorous derivation  of  their leading-order behaviour in the hydrodynamic limit $t\to \infty, \ell\to \infty$ with fixed ratio $t/\ell$. We  first revisit the non-dissipative case~\cite{ares2023lack} in section~\ref{subsec:nodiss} considering the 
quench from the tilted N\'eel state (cf.~\eqref{eq:tiltedNéel}) in the $XX$ chain. 
Specifically, we  provide a thorough analytical proof of the results reported in~\cite{ares2023lack}. 
Our proofs do not rely on Toeplitz structure of the fermionic correlators, unlike the results of Ref.~\cite{ares2023lack}, and 
hence is straightforwardly generalizable to other situations. Then we extend our results accounting  for  the 
effect of gain/loss dissipation in section~\ref{subsec:diss}. 

\subsection{Non-dissipative case}\label{subsec:nodiss}
We start by rewriting Eq.~\eqref{eq:chmom_gamma} as
\begin{equation}
    \label{eq:chmom_log}
    \ln Z_n(\boldsymbol{\alpha},t)=\frac{1}{2}\mathrm{Tr}\left[ \ln\left(\left(\frac{\mathds{1}-\Gamma(t)}{2}\right)\left(\mathds{1}+\prod_{j=1}^n W_j(t)\right)\right)\right].
\end{equation}
The correlation matrix $\Gamma(t)$ (cf.~\eqref{eq:two_point_corr}) can be cast as a block Toeplitz matrix,
\begin{equation}\label{eq:toepl}
    \Gamma_{ll'}(t)=\int_{0}^{2\pi}\frac{dk}{2\pi}\mathcal{G}_t(k)e^{-i k(l-l')}, \qquad l,l'=1,\dots \ell/2. 
\end{equation}
Here the so-called symbol $\mathcal{G}_t(k)$ is a $4\times 4$ matrix
\begin{equation}\label{eq:symbol_t_Néel}
    \mathcal{G}_t(k)=
\left(\begin{array}{cc} 
\mathcal{C}_t(k) & \mathcal{F}_t(k)^\dagger\\
\mathcal{F}_t(k) & -\mathcal{C}_t(-k)^* 
\end{array}\right). 
\end{equation}
For the quench from the tilted N\'eel in the $XX$ chain one has~\cite{ares2023lack} 
\begin{equation}\label{eq:ckt}
    \mathcal{C}_t(k)=
    \left(\begin{array}{cc}
        \cos(2t\varepsilon(k/2)) g_{11}(k, \theta) &
        e^{ik/2}(g_{12}(k,\theta)+i\sin(2t\varepsilon(k/2))g_{11}(k,\theta)) \\
        e^{-ik/2}(g_{12}(k,\theta)-i\sin(2t\varepsilon(k/2))g_{11}(k,\theta)) & -\cos(2t\varepsilon(k/2))g_{11}(k,\theta)
    \end{array}\right),
\end{equation}
and 
\begin{equation}\label{eq:fkt}
    \mathcal{F}_t(k)= 
    \left(\begin{array}{cc}
    i f_{11}(k,\theta)-f_{12}(k,\theta)\sin(2t\varepsilon(k/2)) &
    ie^{ik/2}\cos(2t\varepsilon(k/2))f_{12}(k,\theta)\\
    i e^{-ik/2}\cos(2t\varepsilon(k/2))f_{12}(k,\theta) & 
    -if_{11}(k,\theta)-f_{12}(k,\theta)\sin(2t\varepsilon(k/2))
    \end{array}\right).
\end{equation}
The entries of~\eqref{eq:ckt} and~\eqref{eq:fkt} are given in terms of~\cite{ares2023lack} 
\begin{equation}
\begin{split}
    \label{eq:g11}
    g_{11}(k, \theta)&=-\cos(\theta)-
    \frac{\cos\theta\sin^2\theta(\cos k+\cos^2\theta)}
    {(1+2\cos k \cos^2\theta+\cos^4\theta)},\\
    g_{12}(k, \theta)&=-\frac{\cos(k/2)(1-\cos^4\theta)}
    {1+2\cos k \cos^2\theta+\cos^4\theta},\\
    f_{11}(k,\theta)&=-\frac{\cos\theta\sin^2\theta\sin k}
    {1+2\cos k \cos^2\theta+\cos^4\theta},\\
    f_{12}(k, \theta)&=-\frac{\sin(k/2)\sin^4\theta}
    {1+2\cos k\cos^2\theta+\cos^4\theta}.
    \end{split}
\end{equation}
The main result of this section is that at the leading order in the hydrodynamic limit Eq.~\eqref{eq:chmom_log} gives 
\begin{multline}
    \label{eq:chmom_general}
    \ln Z_n(\boldsymbol{\alpha},t)=\frac{\ell}{4}\int_0^{2\pi} \frac{dk}{2 \pi} \left[ \ln\left(\det\mathcal{Z}_{n,0}^{(\boldsymbol{\alpha})}(k)\right) + \right. \\ \left. \min \left(\frac{2v_kt}{\ell},1\right)\left(\ln\left(\det\mathcal{Z}_{n,\infty}^{(\boldsymbol{\alpha})}(k)\right)-\ln\left(\det\mathcal{Z}_{n,0}^{(\boldsymbol{\alpha})}(k)\right)\right)\right],
\end{multline}
where $\mathcal{Z}_{n,0}^{(\boldsymbol{\alpha})}(k)$ and $\mathcal{Z}_{n,\infty}^{(\boldsymbol{\alpha})}(k)$ are $4\times 4$ matrices (cf.~\eqref{eq:nodiss_t0},~\eqref{eq:nodiss_tinf},~\eqref{eq:angles_tinf}) built, respectively, from the symbol $\gcal_0(k)$ at $t=0$ (cf.~\eqref{eq:toepl}) and 
from the non-oscillating part $\gcal_\infty(k)$ (cf.~\eqref{eq:G_pm0}). In~\eqref{eq:chmom_general} we defined $v(k):=|\varepsilon'(k/2)|$, with $\varepsilon(k)=-\cos(k)$ (cf.~\eqref{eq:XX_diag}). In the following paragraph, we report the \emph{ab initio} derivation of~\eqref{eq:chmom_general}. Notice that Eq.~\eqref{eq:chmom_general} was 
conjecture in Ref.~\cite{ares2023lack} based on robust analytical and numerical evidence. Finally, we remark that the dispersion relation appearing in Eqs.~\eqref{eq:ckt} and~\eqref{eq:fkt} is $\varepsilon(k/2)$, rather than $\varepsilon(k)$. The reason is that the Brillouin
zone of the post-quench Hamiltonian has been halved to rearrange the entries of $\Gamma(t)$ as a block
Toeplitz matrix, taking into account the two-site periodicity of the initial state~\eqref{eq:cat}~\cite{ares2023lack}.

\paragraph{Proof of Eq.~\eqref{eq:chmom_general}.} Let us start the proof by first evaluating the moments $M_{\mathbf{a},\mathbf{b}}$ defined as 
\begin{equation}
    \label{eq:N-mom}
    M_{\mathbf{a},\mathbf{b}}:=\mathrm{Tr}\left[\Gamma(t)^{a_1}N_A^{b_1}...\;\Gamma(t)^{a_n}N_A^{b_n}\right].
\end{equation}
where $N_A$ originates from $W_j(t)$ (cf.~\eqref{eq:chmom_log}). 
The moments $M_{\mathbf{a},\mathbf{b}}$ appear in the Taylor expansion of~\eqref{eq:chmom_log}. 
Analytical knowledge of $M_{\mathbf{a},\mathbf{b}}$ as a function of $\mathbf{a},\mathbf{b},\boldsymbol{\alpha}$, 
in the hydrodynamic limit allows one to obtain $\ln(Z_n(\boldsymbol{\alpha},t))$. 
We exploit the structure for $\Gamma(t)$ in Eq.~\eqref{eq:toepl} 
to write $M_{\mathbf{a},\mathbf{b}}$ as 
\begin{multline}
    \label{eq:stphase_st1}
        M_{\mathbf{a},\mathbf{b}}=\int_0^{2\pi}\frac{dk_1...dk_M}{(2\pi)^M}\prod_{j=1}^M\left(\sum_{x_j=1}^{\ell/2}e^{i(k_j-k_{j+1})x_j}\right)\times\\
    \times\mathrm{Tr}\left[\gcal_t(k_1)...\gcal_t(k_{a_1})n_A^{b_1}...n_A^{b_{n-1}}\gcal_t(k_{a_1+...+a_{n-1}+1})...\gcal_t(k_{a_1+...+a_n})n_A^{b_n}\right],
\end{multline}
where $M:=a_1+...+a_n$ is the total number of quasimomentum variables, $n_A:=\text{diag}(-1,-1,1,1)$, and we identify $k_{M+1}\equiv k_1$. To proceed, we transform the sums in~\eqref{eq:stphase_st1} into integrals, by using the identity
\begin{equation}
    \label{eq:sumtoint}
    \sum_{x=1}^{\ell/2}e^{ikx}=\frac{\ell}{4}\int_{-1}^1 d\xi w(k)e^{i(\ell \xi + \ell +2)k/4},\qquad \text{with}\; w(k):=\frac{k/2}{\sin(k/2)}.
\end{equation}
We then decompose $\gcal_t(k)$ in Eq.~\eqref{eq:symbol_t_Néel} according to the different temporal dependences, i.e.,
\begin{equation}
    \label{eq:G_decomp}
    \gcal_t(k)=U(k)\left(\gcal_\infty(k)+\gcal_+(k)e^{2it\varepsilon(k/2)}+\gcal_-(k)e^{-2it\varepsilon(k/2)}\right)U(k)^{-1},
\end{equation}
where
\begin{align}
    \label{eq:G_inf}
        & \gcal_\infty(k):=\begin{pmatrix}
            0&g_{12}(k,\theta)&-if_{11}(k,\theta)&0\\
            g_{12}(k,\theta)&0&0&if_{11}(k,\theta)\\
            if_{11}(k,\theta)&0&0&-g_{12}(k,\theta)\\
            0&-if_{11}(k,\theta)&-g_{12}(k,\theta)&0
        \end{pmatrix},\\
        \label{eq:G_p}
        & \gcal_+(k):=\frac{1}{2}\begin{pmatrix}
            g_{11}(k,\theta)&g_{11}(k,\theta)&i f_{12}(k,\theta)&-i f_{12}(k,\theta)\\
            -g_{11}(k,\theta)&-g_{11}(k,\theta)&-i f_{12}(k,\theta)&i f_{12}(k,\theta)\\
            i f_{12}(k,\theta)&i f_{12}(k,\theta)&-g_{11}(k,\theta)&g_{11}(k,\theta)\\
            i f_{12}(k,\theta)&i f_{12}(k,\theta)&-g_{11}(k,\theta)&g_{11}(k,\theta)
        \end{pmatrix},\\
        \label{eq:G_m}
        & \gcal_-(k):=\frac{1}{2}\begin{pmatrix}
            g_{11}(k,\theta)&-g_{11}(k,\theta)&-i f_{12}(k,\theta)&-i f_{12}(k,\theta)\\
            g_{11}(k,\theta)&-g_{11}(k,\theta)&-i f_{12}(k,\theta)&-i f_{12}(k,\theta)\\
            -i f_{12}(k,\theta)&i f_{12}(k,\theta)&-g_{11}(k,\theta)&-g_{11}(k,\theta)\\
            i f_{12}(k,\theta)&-i f_{12}(k,\theta)&g_{11}(k,\theta)&g_{11}(k,\theta)
        \end{pmatrix},\\
        \label{eq:G_pm0}
        & \gcal_0(k):=\gcal_\infty+\gcal_++\gcal_-. 
\end{align}
In~\eqref{eq:G_decomp} we have factored out the matrix $U(k)$ 
\begin{equation}
    \label{eq:U}
    U(k):=\text{diag}(1,e^{-i\frac{k}{2}},1,e^{-i\frac{k}{2}}),
\end{equation}
and its inverse, which cancel out when we take the trace in~\eqref{eq:stphase_st1}. Now, after substituting  the 
decomposition~\eqref{eq:G_decomp} in~\eqref{eq:stphase_st1}, one obtains $3^M$  terms. 
Each term is the trace of a string of powers of $\gcal_\pm(k),\gcal_\infty(k),n_A$ multiplied by a time-oscillating phase. We observe that most of the traces of these strings are zero. Indeed, it is not hard to explicitly check that, because of the symmetries of $\gcal_\pm(k)$ and $\gcal_\infty(k)$, we have:
\begin{itemize}
    \item $\gcal_+(k)P\gcal_+(k')=\gcal_-(k)P\gcal_-(k')=0$, where $P$ is any product of matrices $\gcal_\infty(k)$ and $n_A$;
    \item $\mathrm{Tr}\left[\gcal_\pm(k) P\right]=0$, with $P$ as before.
\end{itemize}
Thus, the only strings of matrices that give a nonzero contribution to the moments~\eqref{eq:stphase_st1} are those containing substrings of alternated $\gcal_+(k)$ and $\gcal_-(k)$, possibly interspersed with factors $\gcal_\infty(k)$ and $n_A$.

Let us  now perform the integration in the $2M-2$ variables $k_2,...,k_M,\xi_2,...,\xi_M$ in Eq.~\eqref{eq:stphase_st1}, applying the stationary phase approximation. In the limit $\ell\to\infty$, the stationary phase states that 
\begin{equation}
  \label{eq:stphase}
  \int_\mathcal{D} d^Nx\; g(\mathbf{x})e^{i\ell f(\mathbf{x})} = \sum_j \left(\frac{2\pi}{\ell}\right)^{\frac{N}{2}}g(\mathbf{x}_j)|\det H(\mathbf{x}_j)|^{-\frac{1}{2}}e^{i\ell f(\mathbf{x}_j)+\frac{i\pi\sigma(\mathbf{x}_j)}{4}}+o(\ell^{-\frac{N}{2}}),
\end{equation}
where $\mathbf{x}_j\in\mathcal{D}$ are the stationary points of $f(\mathbf{x})$, i.e., such that $\nabla f(\mathbf{x}_j)=0$, $\mathcal{D}$ is the integration domain, $H$ is the Hessian matrix of $f$ and $\sigma$ its signature (precisely, the difference between the number of positive and negative eigenvalues). Let us study the stationarity conditions for the generic term in~\eqref{eq:stphase_st1}. The derivatives of the phase with respect to the $\xi_j$ variables  give the conditions $k_j=k_{j+1}, \;j=2,...,M$, which imply $k_j=k_1 \; \forall j$.  By taking the derivative of the phase with respect to the $k_j$ variables, we obtain 
\begin{align}
    \label{eq:stationarity_k}
   & \frac{1}{4}(\xi_j-\xi_{j-1})\pm\frac{t}{\ell}\varepsilon'_{k_j/2}=0,\\
    \label{eq:stationarity_k-1}
   & \frac{1}{4}(\xi_j-\xi_{j_1})=0, 
\end{align}
with $j=2,\dots, M$. Precisely, the stationarity condition with respect to $k_j$ 
gives~\eqref{eq:stationarity_k} if a term $\gcal_{\pm}(k_j)$ is present in the string of operators appearing in~\eqref{eq:stphase_st1}. The $\pm$ in~\eqref{eq:stationarity_k} is the same as in $\gcal_\pm(k_j)$. If $k_j$ is not present in the string, 
the stationarity condition gives~\eqref{eq:stationarity_k-1}. 
 Keeping in mind that only  strings with alternate occurrences of $\gcal_\pm(k)$ can appear in~\eqref{eq:stphase_st1}, the stationarity conditions~\eqref{eq:stationarity_k} imply that either $\xi_j\in\{\xi_1,\xi_1+4\varepsilon'(k_1/2)t/\ell\}\; \forall j \in [2,M]$ or $\xi_j\in\{\xi_1,\xi_1-4\varepsilon'(k_1/2)t/\ell\}\; \forall j \in [2,M]$, except for the only string containing only $\gcal_\infty(k)$ and $n_A$, which gives $\xi_j=\xi_1 \;\forall j$. 
 
 Since any stationary point  has to be inside the integration domain, one must have  $-1<\xi_j<1\;\forall j \in [2,M]$. This condition puts a constraint on $\xi_1$, except for the string without any $\gcal_\pm(k)$.  Thus, by integrating over $\xi_1$, the terms originating from strings with at least a $\gcal_\pm(k)$ yield a factor $\max\left(0,2-4|\varepsilon'(k_1/2)|\frac{t}{\ell}\right)=2\left(1-\min\left(2|\varepsilon'(k_1/2)|t/\ell,1\right)\right)$, whereas  the strings with only $\gcal_\infty$ and $n_A$ give a factor $2$. We can also compute the determinant of the Hessian matrix evaluated at the stationary points (cf.~\eqref{eq:stphase}), which is $4^{2-2M}$, and its signature, which is simply $0$. Substituting everything back into Eq.~\eqref{eq:stphase_st1}, and evaluating the functions in the stationary point, we obtain
\begin{multline}
    \label{eq:stphase_final}
        M_{\mathbf{a},\mathbf{b}}=\frac{\ell}{2}\int_0^{2\pi}\frac{dk}{2\pi}\left[\mathrm{Tr}\left[\gcal_\infty(k)^{a_1}n_A^{b_1}...\gcal_\infty(k)^{a_n}n_A^{b_n}\right]+\left(1-\min\left(\frac{2v_k t}{\ell},1\right)\right)\times\right.\\
        \left.\times\left(\mathrm{Tr}\left[\gcal_0(k)^{a_1}n_A^{b_1}...\gcal_0(k)^{a_n}n_A^{b_n}\right]-\mathrm{Tr}\left[\gcal_\infty(k)^{a_1}n_A^{b_1}...\gcal_\infty(k)^{a_n}n_A^{b_n}\right]\right)\right],
\end{multline}
where $\gcal_0(k)=\gcal_\infty(k)+\gcal_+(k)+\gcal_-(k)$ (cf.~\eqref{eq:G_pm0}). 

Notice that now the terms inside the trace 
in~\eqref{eq:stphase_final} do not depend on time, unlike in~\eqref{eq:stphase_st1}. The only 
time dependence is in the kinematic factor $\min(2v_kt/\ell,1)$. Notice also that at $t=0$ and $t\to\infty$ 
only the terms with $\gcal_{0}$ and $\gcal_\infty$ remain, respectively. We find it useful to rewrite Eq.~\eqref{eq:stphase_final} as 
\begin{multline}
    \label{eq:NA-mom_final}
        M_{\mathbf{a},\mathbf{b}}=\frac{\ell}{2}\int_0^{2\pi}\frac{dk}{2\pi}\left[\mathrm{Tr}\left[\gcal_0(k)^{a_1}n_A^{b_1}...\gcal_0(k)^{a_n}n_A^{b_n}\right]-\min\left(\frac{2v_k t}{\ell},1\right)\times\right.\\
        \left.\times\left(\mathrm{Tr}\left[\gcal_0(k)^{a_1}n_A^{b_1}...\gcal_0(k)^{a_n}n_A^{b_n}\right]-\mathrm{Tr}\left[\gcal_\infty(k)^{a_1}n_A^{b_1}...\gcal_\infty(k)^{a_n}n_A^{b_n}\right]\right)\right].
\end{multline}
Now, let us  define $\ln(\mathcal{Z}_{n,0}^{(\boldsymbol{\alpha})}(k))$ by replacing  $\Gamma(t)\to\gcal_0(k)$ and $N_A\to n_a$ in~\eqref{eq:chmom_log}. Analogously, we define $\ln(\mathcal{Z}_{n,\infty}^{(\boldsymbol{\alpha})}(k))$ by replacing $\Gamma(t)\to\gcal_\infty(k)$ and $N_A\to n_A$ in~\eqref{eq:chmom_log}. Thus, analiticity of~\eqref{eq:chmom_log} as a function of $\Gamma$ and $N_A$ implies that the leading order of the charged 
moments in the hydrodynamic limit is Eq.~\eqref{eq:chmom_general}. Specifically, to obtain~\eqref{eq:chmom_general} one can Taylor expand~\eqref{eq:chmom_log},  
apply~\eqref{eq:NA-mom_final} to each term of the series, and resum the obtained expression. 

To proceed, one has to determine $\ln(\det\mathcal{Z}^{\boldsymbol{\alpha}}_{n,0})$ and 
$\ln(\det\mathcal{Z}^{\boldsymbol{\alpha}}_{n,\infty})$. Crucially, this involves 
computing functions of $4\times4$ matrices, unlike~\eqref{eq:chmom_log}, which requires dealing 
with $2\ell\times 2\ell$ matrices. 
The derivation is quite technical and we report it in Appendix~\ref{app:proof_nodiss}. The final 
result for $\ln( \det \mathcal{Z}_{n,0}^{(\boldsymbol{\alpha})})$ reads
\begin{equation}
    \label{eq:nodiss_t0}
    \ln \left( \det \mathcal{Z}_{n,0}^{(\boldsymbol{\alpha})} \right)= \ln \prod_{j=1}^n \left(1-m_N^2(k,\theta) \sin^2(\alpha_{j,j+1})\right),
\end{equation}
where $m_N(k,\theta)=\sqrt{f_{11}(k,\theta)^2+f_{12}(k,\theta)^2}$ and $f_{ij}$ are defined 
in~\eqref{eq:g11}. Moreover, one obtains  $\ln( \det \mathcal{Z}_{n,\infty}^{(\boldsymbol{\alpha})})$ as 
\begin{equation}
    \label{eq:nodiss_tinf}
    \ln \left( \det \mathcal{Z}_{n,\infty}^{(\boldsymbol{\alpha})} \right)= 2 \ln \left( h_n(n_+(k,\theta))h_n(n_-(k,\theta))-f_{11}(k,\theta)^2 \tilde{h}_n(\boldsymbol{\alpha},k,\theta) \right),
\end{equation}
with 
\begin{align}
\label{eq:npm}
& n_\pm(k,\theta):=\frac{1}{2}(1+g_{12}(k,\theta)\pm f_{11}(k,\theta))\\
\label{eq:hn}
& h_n(x):=x^n+(1-x)^n
\end{align}
and 
\begin{multline}
    \label{eq:angles_tinf}
        \tilde{h}_n(\boldsymbol{\alpha},k,\theta):=\sum_{j=1}^{\lfloor \frac{n}{2}\rfloor} \frac{f_{11}(k,\theta)^{2j-2}(n_+(k,\theta)+n_-(k,\theta)-2n_+(k,\theta)n_-(k,\theta))^{n-2j}}{2^{n-2}}\times\\
        \times \sum_{1\leq p_1<p_2<...<p_{2j}\leq n} \sin^2(\alpha_{p_1}-\alpha_{p_2}+...-\alpha_{p_{2j}}).
\end{multline}
Again, Eq.~\eqref{eq:nodiss_t0} and Eq.~\eqref{eq:nodiss_tinf} were conjectured in 
Ref.~\cite{ares2023lack}. 

\subsection{Dissipative case}\label{subsec:diss}

Let us now focus on the time evolution of the charged moments~\eqref{eq:Znalpha} 
in the presence of fermionic gain and loss processes (see section~\ref{sec:models}). 

In the presence of gain and loss dissipation, the symbol of $\Gamma(t)$ in Eq.~\eqref{eq:symbol_t_Néel} must be modified. In particular, the correlations $\mathcal{C}_t$ and $\mathcal{F}_t$ (cf.~\eqref{eq:symbol_t_Néel}) are affected differently by the dissipation. A straightforward calculation using the formalism of Ref.~\cite{prosen2008third} shows that in the 
presence of gain and loss one has that 
\begin{equation}
    \label{eq:G_diss}
    \gcal'_t(k)=e^{-(\gamma_+ + \gamma_-)t}\gcal_t(k)+(2n_\infty-1)(1-e^{-(\gamma_+ + \gamma_-)t})\sigma^z\otimes\mathds{1},
\end{equation}
where $n_{\infty}:=\gamma_+/(\gamma_++\gamma_-)$, with $\gamma_\pm$ the dissipation rates. 
In~\eqref{eq:G_diss} the symbol $\otimes$ denotes the Kronecker product. The term $\sigma^z\otimes\mathds{1}$ implies  
that while the correlator $\mathcal{C}_t$ 
is shifted and rescaled due to the dissipation, $\mathcal{F}_t$ is only rescaled. 
In order to compute the entanglement asymmetry, we start by studying the time evolution of the charged moments in the weakly-dissipative hydrodynamic limit $t, \ell \to \infty$, $\gamma_+,\gamma_- \to 0$ with fixed $t/\ell$, $\gamma_+ t$, $\gamma_-t$. Similar to the non-dissipative case, we anticipate that  the leading order contribution in the weakly-dissipative hydrodynamic limit of Eq.~\eqref{eq:chmom_log} is given as 
\begin{multline}
    \label{eq:chmom_general_diss}
    \ln Z_n(\boldsymbol{\alpha},t)=\frac{\ell}{4}\int_0^{2\pi} \frac{dk}{2 \pi} \left[ \ln\left(\det\mathcal{Z'}_{n,0}^{(\boldsymbol{\alpha})}(k,t)\right) +\right.\\ \left. \min \left(\frac{2v_kt}{\ell},1\right)\left(\ln\left(\det\mathcal{Z'}_{n,\infty}^{(\boldsymbol{\alpha})}(k,t)\right)-\ln\left(\det\mathcal{Z'}_{n,0}^{(\boldsymbol{\alpha})}(k,t)\right)\right)\right], 
\end{multline}
which is one of the main results of this section. 
Now $\mathcal{Z'}_{n,0}^{(\boldsymbol{\alpha})}(k,t)$ and $\mathcal{Z'}_{n,\infty}^{(\boldsymbol{\alpha})}(k,t)$ are the same as 
$\mathcal{Z}^{\boldsymbol{\alpha}}_{n,0}(k,t)$ and $\mathcal{Z}^{\boldsymbol{\alpha}}_{n,\infty}(k,t)$ (cf.~\eqref{eq:chmom_general}), although they are built from the symbol $\gcal'_t(k)$ (cf.~\eqref{eq:G_diss}). Notice that unlike the non-dissipative case, there is a residual time dependence in~\eqref{eq:chmom_general_diss} due to the terms $\exp({-\left(\gamma_++\gamma_-\right)t})$, which, however, are constant in the weakly-dissipative hydrodynamic limit. In the following paragraph, we derive Eq.~\eqref{eq:chmom_general_diss}. We anticipate that, in the presence of gain and loss, we are able to analytically compute $Z_n(\boldsymbol{\alpha},t)$ only in some particular cases. 

\paragraph{Proof of Eq.~\eqref{eq:chmom_general_diss}:} The derivation of Eq.~\eqref{eq:chmom_general} can be adapted to the dissipative case  if we split the symbol $\gcal'_t(k)$ (see Eq.~\eqref{eq:G_diss}) as
\begin{equation}
    \label{eq:G'_decomp}
    \gcal'_t(k)=U(k)\left(\gcal'_\infty(k,t)+\gcal'_+(k,t)e^{2it\varepsilon(k/2)}+\gcal'_-(k,t)e^{-2it\varepsilon(k/2)}\right)U^{-1}(k),
\end{equation}
with
\begin{align}
    \label{eq:G'_pm0}
        &\gcal'_\infty(k,t):=e^{-(\gamma_+ + \gamma_-)t}\gcal_\infty(k)+(2n_\infty-1)(1-e^{-(\gamma_+ + \gamma_-)t})\sigma^z\otimes\mathds{1},\\
            \label{eq:G'_pm0-1}
        &\gcal'_+(k,t):=\gcal_+(k)e^{-(\gamma_+ + \gamma_-)t},\\
            \label{eq:G'_pm0-2}
        &\gcal'_-(k,t):=\gcal_-(k)e^{-(\gamma_+ + \gamma_-)t},\\
            \label{eq:G'_pm0-3}
         &\gcal'_0(k,t):=\gcal'_\infty(k,t)+\gcal'_+(k,t)+\gcal'_-(k,t).
\end{align}
Here $\varepsilon(k)$ is the energy dispersion~$\eqref{eq:XX_diag}$, and $\gcal_\pm(k)$ 
are defined in~\eqref{eq:G_pm0}. Notice that in defining $\gcal'_0$ we performed the $t\to0$ limit 
in~\eqref{eq:G'_decomp} keeping the terms $\exp(-(\gamma_++\gamma_-)t)$ because they are constant 
in the weakly-dissipative hydrodynamic limit. 
Clearly, Eq.~\eqref{eq:chmom_general_diss} is more involved than in the non-dissipative case. Still, the derivation of~\eqref{eq:chmom_general_diss} proceeds as for 
the non-dissipative (see section~\ref{subsec:nodiss}). Precisely, we can still define the moments $M_{\mathbf{a},\mathbf{b}}$ as in~\eqref{eq:N-mom}. In~\eqref{eq:stphase_st1}, one has to replace $\gcal_t$ with $\gcal'_t$ (cf.~\eqref{eq:G'_decomp}). The conditions on the sequences of $\gcal'$ yielding zero trace are the same as for the $\gcal$, thus Eq.~\eqref{eq:NA-mom_final} remains the same after replacing $\gcal_\pm,\gcal_0,\gcal_\infty$ with $\gcal'_\pm,\gcal'_0,\gcal'_\infty$ (cf.~\eqref{eq:G'_pm0}-\eqref{eq:G'_pm0-3}). Finally, this 
allows us to define $\mathcal{Z'}_{n,0}^{(\boldsymbol{\alpha})}(k,t)$ and $\mathcal{Z'}_{n,\infty}^{(\boldsymbol{\alpha})}(k,t)$,  obtaining~\eqref{eq:chmom_general_diss}. 
Now, by using~\eqref{eq:chmom_general_diss}, one can obtain $\ln(Z_n(\boldsymbol{\alpha},t))$ for any $n$ in the 
weakly-dissipative hydrodynamic limit. The computation requires diagonalizing  $4\times 4$ matrices. 
Crucially, to understand the behavior of the asymmetry $\Delta S_A^{(n)}$ (cf.~\eqref{eq:def_renyi}), 
one has to know the analytic dependence on $n$ of the charged moments. Unfortunately, we are not 
able to find a closed-form expression for $\ln(\det\mathcal{Z}'^{(\boldsymbol{\alpha})}_{n,0})$ and $\ln(\det\mathcal{Z}'^{(\boldsymbol{\alpha})}_{n,\infty})$ for generic $\gamma_\pm$. 

Analytic manipulations 
are somewhat easier for balanced gain and losses, i.e., for $\gamma_+=\gamma_-$, which we now discuss. 
We have $n_\infty=1/2$ and Eq.~\eqref{eq:G_diss} becomes $\gcal'_t(k)=\lambda(t)\gcal_t(k)$, where $\lambda(t):=e^{-(\gamma_++\gamma_-)t}$. Again, even for balanced gains and losses, although one can numerically obtain $\gcal_{n,0}$ for 
any integer $n$, a closed-form analytic expression cannot be obtained. 
We refer to Appendix~\ref{app:diss} for more details. Nevertheless, we obtain a closed formula for 
$\ln ( \det \mathcal{Z'}_{n,\infty}^{(\boldsymbol{\alpha})} )$ 
for generic $n$. To do that, one has to multiply $f_{11}(k,\theta)$ and $g_{12}(k,\theta)$ in Eq.~\eqref{eq:g11} by $\lambda(t)$ in Eq.~\eqref{eq:nodiss_tinf}. The derivation is  discussed in Appendix~\ref{app:diss}. The final result reads as 
\begin{equation}
    \label{eq:zn_gpgm_tinf}
    \ln \left( \det \mathcal{Z'}_{n,\infty}^{(\boldsymbol{\alpha})} \right)= 2 \ln \left( h_n(n'_+(k,\theta,t))h_n(n'_-(k,\theta,t))-\lambda(t)^2 f_{11}(k,\theta)^2 \tilde{h}'_n(\boldsymbol{\alpha},k,\theta,t) \right),
\end{equation}
where $n'_\pm (k,\theta,t):=(1+\lambda(t)g_{12}(k,\theta)\pm\lambda(t)f_{11}(k,\theta))/2$ and $\tilde{h}'_n(\boldsymbol{\alpha},k,\theta,t)$ is obtained by replacing $f_{11}(k,\theta)\rightarrow \lambda(t)f_{11}(k,\theta)$ and $n_\pm(k,\theta) \rightarrow n'_\pm (k,\theta,t)$ in Eq.~\eqref{eq:angles_tinf}.  The difficult term to evaluate for generic $n$ is the short-time one $\ln\left(\det \mathcal{Z'}_{2,0}^{(\boldsymbol{\alpha})}\right)$, but we are, at least, able to find a closed formula for $n=2$, reading 
\begin{equation}
    \label{eq:z2_gpgm_t0}
    \ln\left(\det \mathcal{Z'}_{2,0}^{(\boldsymbol{\alpha})}\right)=2 \ln \left(\frac{(1+\lambda^2)^2}{4} -\lambda^2 m_N(k,\theta)^2\sin^2(\alpha_{1,2})\right).
\end{equation}

If we relax the condition $\gamma_+=\gamma_-$, the formulas become even more cumbersome, and we report them only for small $n$. For $n=2$ we find 
\begin{equation}
\begin{split} \label{eq:z2_generic_t0}
       & \ln\left(\det \mathcal{Z'}_{2,0}^{(\boldsymbol{\alpha})}\right)=\ln \left[\lambda(t)^2(m_N(k,\theta)^2-1)s^2((\lambda(t) -1)^3 s^2+(\lambda(t)-1)(\lambda(t)^2+1))^2 +\right. \\ &\left.+\left(\frac{(1+\lambda(t)^2)^2+2(\lambda(t)-1)^2(1-2\lambda(t)^2 m_N(k,\theta)^2+3\lambda(t)^2)s^2+(\lambda(t)-1)^4s^4}{4} \right.\right.\\
       &\left. \left.-\lambda(t)^2 m_N(k,\theta)^2\sin^2(\alpha_{1,2})\right)^2\right].
\end{split}
\end{equation}
Here we introduced $s:=2n_\infty-1=(\gamma_+-\gamma_-)/(\gamma_++\gamma_-)$. Similarly, we can derive  the long-time term $\ln \left( \det \mathcal{Z'}_{n,\infty}^{(\boldsymbol{\alpha})} \right)$ only for small values of $n$. We give some detail in Appendix~\ref{app:diss}. For instance, for $n=2$ we obtain 
\begin{equation}
  \label{eq:z2_generic_tinf}
  \ln\left(\det \mathcal{Z'}_{2,\infty}^{(\boldsymbol{\alpha})}\right)=2 \ln \left(h_2(n'_+(k,\theta,t))h_2(n'_-(k,\theta,t))-\lambda^2(t) f_{11}(k,t)^2\tilde{h}'_2(\boldsymbol{\alpha},k,\theta,t) \right),
\end{equation}
where $\tilde{h}'_n(\boldsymbol{\alpha},k,\theta,t)$ is obtained by substituting $f_{11}(k,\theta)\rightarrow \lambda(t)f_{11}(k,\theta)$ and $n_\pm(k,\theta) \rightarrow n'_\pm (k,\theta,t)$ in the definition~\eqref{eq:angles_tinf} of $\tilde{h}_n(\boldsymbol{\alpha},k,\theta)$. Since $\gamma_+\neq \gamma_-$, one has \\ $n'_\pm (k,\theta,t)=\left(1+\lambda(t)g_{12}(k,\theta)\pm\sqrt{\lambda(t)^2f_{11}(k,\theta)^2+s^2(\lambda(t)-1)^2}\right)/2$.
The results above allow us to obtain an analytic expression for the leading-order term of $\ln Z_2(\boldsymbol{\alpha},t)$ for generic $\gamma_\pm$, by plugging Eqs.~\eqref{eq:z2_generic_t0} and~\eqref{eq:z2_generic_tinf} in  Eq.~\eqref{eq:chmom_general_diss}. We should stress that the 
case with $n=2$ is sufficient to probe the Mpemba effect, as it has been recently shown~\cite{joshi2024observing}.

\section{Generalized quasiparticle picture for the charged moments}\label{sec:chmom_quasipart}

Both Eqs.~\eqref{eq:chmom_general} and~\eqref{eq:chmom_general_diss} have a form that is reminiscent of  the quasiparticle picture of the entanglement dynamics~\cite{calabrese2005evolution,fagotti2008evolution,alba2017entanglement,calabrese2018entanglement}. The main ingredient of the quasiparticle picture is the presence of entangled multiplets of excitations, which are formed after the quench. In the most basic incarnation, the multiplets are formed by two quasiparticles, i.e., by  entangled pairs of quasiparticles. 
The entangled quasiparticles travel ballistically, as free excitations. 
The entanglement entropy between two complementary subsystems is proportional to the number 
of entangled pairs shared between the intervals. The entanglement content shared by each pair is related to the 
thermodynamic entropy  of the GGE describing the post-quench steady state~\cite{fagotti2008evolution,alba2017entanglement}. 
For free-fermion and free-boson models, the quasiparticle picture describes the 
dynamics of both von Neumann and R\'enyi entropies, and the associated mutual 
information~\cite{alba2018entanglement}. The quasiparticle picture can be also used to 
describe the dynamics of the logarithmic negativity~\cite{alba2019quantum}. The dynamics 
of quantum information after quenches from inhomogeneous initial states has been addressed~\cite{bertini2018entanglementevolution,alba2018entanglementand,mestyan2020molecular,alba2021generalized}. 
Very recently, the quasiparticle picture has been adapted to describe entanglement dynamics in 
systems described by quadratic fermionic and bosonic Lindblad master equations~\cite{alba2021spreading,carollo2022dissipative,alba2022hydrodynamics,alba2022logarithmic}. 
The generalization of the quasiparticle picture for interacting integrable models was 
conjectured in Ref.~\cite{alba2017entanglement} for quenches producing entangled pairs only. Interestingly, 
while a quasiparticle picture works for the von Neumann entropy~\cite{alba2017entanglement,klobas2021entanglement,klobas2021exact}, it has been shown in Ref.~\cite{bertini2022growth} that in the presence of interactions it fails for 
the R\'enyi entropies, although a hydrodynamic 
description of the growth of the R\'enyi entropies is possible. The steady-state R\'enyi entropies 
have been obtained in~\cite{alba2017quench,alba2017renyi,mestyan2018renyi} also for interacting systems. 
Very recently, the validity of the quasiparticle picture for two-dimensional free systems has been explored in Refs.~\cite{yamashika2023time,gibbins2023quench}. 
Remarkably, the quasiparticle picture, 
at least for free models, can be extended to quantum quenches producing entangled multiplets, i.e., 
going beyond the entangled pair paradigm~\cite{bastianello2018spreading, caceffo2023negative,bertini2018entanglement}. 

Here we show that the quasiparticle picture for entangled multiplets~\cite{bastianello2018spreading,caceffo2023negative} provides the natural 
framework to describe the dynamics of the charged moments in free-fermion systems. Notice that although 
some elements of the quasiparticle picture were presented already in Refs.~\cite{ares2023entanglement,rylands2023microscopic,murciano2023entanglement}, a full-fledged 
quasiparticle picture was still lacking. Specifically, even in the case of quenches producing entangled pairs, a 
framework to 
determine the contribution of the quasiparticles to the charged moments and the asymmetry for generic initial states was not completely specified. Therefore, we conjecture a formula for $\ln(Z_{n}(\boldsymbol{\alpha},t))$, which allows to describe their evolution in the hydrodynamic limit. We describe our framework for the non-dissipative case, showing that it is consistent with the  results  of Sec.~\ref{subsec:nodiss} (and also with the results of~\cite{ares2023lack}). For the dissipative case, we do not have a well-established framework, even though the results of Ref.~\cite{alba2021generalized,alba2021spreading,carollo2022dissipative}  suggest that it should be 
possible to generalize our approach to that case.

\subsection{Entangled multiplets and von Neumann entropy}
\label{sec:quasi-ent}

Let us briefly review the quasiparticle picture for the von Neumann entropy $S_A(\rho_A)$ in the presence of entangled multiparticle excitations. 
As anticipated, this generalized quasiparticle picture explains the spreading of quantum correlations as a consequence of the propagation of entangled \textit{multiplets} of quasiparticles. 
Let us assume that the Hamiltonian of the system can be recast in the form~\eqref{eq:gen-quasipart_1}, with multiplets formed by $m$ quasiparticles.  
Let us assume that in the basis of the post-quench fermionic modes $d_j(k)$ (cf~\eqref{eq:gen-quasipart_1}), the two-point correlation function on the initial state is block-diagonal in the quasimomentum space, i.e., if we define $D(k)^\dagger:=(d_1(k),...,d_m(k),d_1^\dagger(k),...,d_m^\dagger(k))$, we have 
    \begin{equation}
        \label{eq:gen-quasipart_2}
        \langle \Psi(0)|D(k)D(h)^\dagger|\Psi(0)\rangle =\delta(k-h) \langle \Psi(0)|D(k)D(k)^\dagger|\Psi(0)\rangle:=\delta(k-h)\tilde{G}(k). 
    \end{equation}
    The correlator $G(k):=2\tilde{G}(k)-\mathds{1}$  contains the correlations among the various species of particles constituting the multiplet with quasimomentum $k$, and it can be usefully decomposed into $m\times m$ blocks as
    \begin{equation}
        \label{eq:gen-quasipart_3}
        G(k):=\begin{pmatrix}
            C(k)&F(k)^\dagger\\
            F(k)&-C(k)^T
        \end{pmatrix}, 
    \end{equation}
where $C(k):=2\langle d_i^\dagger(k) d_j(k)\rangle-\mathds{1}$, with $i,j=1,\dots m$ labelling the different species, 
and $F(k):=2\langle d_i(k)d_j(k)\rangle$. 
    
Let us now focus on the dynamics of the entangled multiplets.  
One has to be careful in identifying the propagation velocity, that in general 
is \textit{not}  $v_j(k)=d\varepsilon_j(k)/dk$, with $k$ the quasimomentum in the folded Brillouin zone $\mathcal{B}=[0,2\pi/m]$. This is due to the fact that  the quasimomentum $k\in[0,2\pi/m]$ of the quasiparticles of 
species $j$ is a nontrivial function $k(k')$ of the quasimomentum $k'\in[0,2\pi]$ before the folding. 
In general, $k$ is related to $k'$ by a shift and a sign change. 
On the other hand, the velocity of the quasiparticles 
has to be obtained from the dispersion $\varepsilon(k')$ before the folding, because $\varepsilon(k')$ 
is the dispersion of the physical excitations of the system. Thus, we have 
$v_j(k)=d\varepsilon_j(k')/dk'= d\varepsilon_j(k)/dk \cdot dk/dk'$, where $dk/dk'$ accounts for the 
minus sign originating from the folding. 

Finally, the contribution $s_A(x,t,k)$ to the von Neumann entropy 
of a  subsystem $A$ at time $t$ originating from a multiplet produced initially at position $x$  is conjectured 
to be~\cite{bertini2018entanglementevolution,bastianello2018spreading} 
\begin{equation}
    \label{eq:gen-quasipart_sk}
    s_A(x,t,k):=-\text{Tr}\left[\frac{G_{\mathcal{A}(x,t)}(k)+\mathds{1}}{2}\ln\frac{G_{\mathcal{A}(x,t)}(k)+\mathds{1}}{2}\right],
\end{equation}
where $G_{\mathcal{A}(x,t)}$ is the restriction of the matrix $G(k)$ (cf.~\eqref{eq:gen-quasipart_3}) to the indices describing the subset $\mathcal{A}(x,t)$ of quasiparticles that are contained in the subsystem $A$ at time $t$. The R\'enyi entropies can be found in a similar way. 
To understand Eq.~\eqref{eq:gen-quasipart_sk}, it should be noticed that the correlator in quasimomentum-space (cf.~\eqref{eq:gen-quasipart_2}) and the real-space correlator $(\Gamma+\mathds{1})/2$, with $\Gamma$ defined in~\eqref{eq:two_point_corr}, have the same spectrum, because the change of basis to go to momentum space is unitary (the canonical anticommutation rules are preserved). 
Clearly, this does not hold for the spectrum of the correlator restricted to a subsystem $A$, because the 
bipartition breaks the translation symmetry. This renders the computation of entanglement-related quantities challenging. However, one can imagine of dividing the chain into ``mesoscopic'' cells of lengtt $dx$, such that $a\ll dx \ll |A|$, where $a$ is the lattice spacing. Each cell is characterized by a density matrix $\rho_x$, which, since the system is in a Gaussian state, is fully described by the two-point function $G_x$, whose spectrum, since $dx\gg a$, will be approximately the same as that of full system, i.e., $\bigoplus_{k\in \mathcal{B}} G(k)$. Thus, we can write $\rho_x\approx\bigotimes_{k\in\mathcal{B}}\rho_{x,k}$, with the state of the whole chain $\rho=\bigotimes_{x}\rho_{x}$, and $\rho_{x,k}$ the density matrix identified by $G_x(k)$. To connect with the standard quasiparticle picture, we  view each cell as a source of entangled multiplets of quasiparticles, the quasiparticles being the single-particle eigenmodes $d_1^\dagger(k), ...,d_m^\dagger(k)$ of the Hamiltonian. The correlations between them are encoded in  $G_x(k)$. The multiplets spread because of the ballistic motion of the quasiparticles. Suggestively, we could associate to each multiplet $k$ produced in the cell at $x$ a non-local correlation matrix $G_{x,k,t}$. This matrix describes the global state of the $m$  cells at positions $x+v_j(k)t\rangle\}_{j=1,...,m}$. The only nonzero entries of $G_{x,k,t}$ are the same as 
the entries of $G_{x,k}$ with the caveat that the mode $d_j^\dagger$ is that of the cell at $x+v_j(k)t$, meaning that 
the different modes are in different cells. 
So $G_{x,k,t}$ describes the way in which different cells are correlated with each other by the modes originating from the cell at $x$. Since the state is 
Gaussian, one could associated to $G_{x,k,t}$ a density matrix $\rho_{x,k,t}$ describing the global state of the 
cells. 
 The state of the whole chain is, then, $\rho_{t}=\bigotimes_{x}\rho_{x,t}\approx\bigotimes_{x}\bigotimes_{k\in\mathcal{B}}\rho_{x,k,t}$. Now the state of a subsystem $A$ is obtained by tracing over the cells that are not in $A$ in each $\rho_{x,k,t}$, which corresponds to tracing over the modes outside of $A$ in the matrix $G_{x,k,t}$. Finally, from the state of $A$ one can compute the entanglement entropy, and the result is~\eqref{eq:gen-quasipart_sk}.

Notice also that in the case $m=2$ this prescription yields the well-known formula for the case of entangled pairs (see e.g.~\cite{fagotti2008evolution}), as shown in~\cite{caceffo2023negative}.
It has been checked in~\cite{bastianello2018spreading} that our procedure gives the correct hydrodynamic limit for the von Neumann and R\'enyi entropies, and also for the  tripartite mutual information in~\cite{caceffo2023negative}. Moreover, this procedure allows to obtain the dynamics of the von Neumann entropy 
starting from inhomogeneous initial states~\cite{bertini2018entanglementevolution}.

\subsection{Entangled multiplets and charged moments}
\label{sec:mult-quasi}

Having explained the quasiparticle picture for the von Neumann entropy in the 
presence of entangled multiplets, we now show how it can be adapted to describe the charged moments (cf.~Eq.~\eqref{eq:chmom_gamma}). Inspired by Eq.~\eqref{eq:chmom_log}, since $\ln Z_n(\boldsymbol{\alpha},t)$ is a function of the real-space correlator $\Gamma$ restricted to the considered subsystem, we can  generalize the argument leading to~\eqref{eq:gen-quasipart_sk}.  We conjecture that the leading order behaviour of $\ln Z_n(\boldsymbol{\alpha},t)$, in the hydrodynamic limit and for any subsystem $A$, is obtained as a sum of the contributions $z_n(x,t,k)$ from each multiplet, reading
\begin{equation}
    \label{eq:chmom_quasipart}
    z_n(x,t,k):=\frac{1}{2}\mathrm{Tr}\left[ \ln\left(\left(\frac{\mathds{1}-G_{\mathcal{A}(x,t)}(k)}{2}\right)^n\left(\mathds{1}+\prod_{j=1}^n w_j^{\mathcal{A}}(x,t,k)\right)\right)\right],
\end{equation}
where $G_{\mathcal{A}(x,t)}$ is the same as in~\eqref{eq:gen-quasipart_sk}. 
%{\color{red}In the same spirit of~\eqref{eq:gen-quasipart_sk}, $z_n(x,t,k)$ is the charged moment computed from the density matrix $\mathrm{Tr}_{out}\rho_{-k,k}$, in which we trace on the modes that are outside of $A$. }
Here, 
we defined $w_j^{\mathcal {A}}(x,t,k)$ as 
\begin{equation}\label{eq:wj}
    w_j^{\mathcal{A}}(x,t,k):=\frac{\mathds{1}+G_{\mathcal{A}(x,t)}(k)}{\mathds{1}-G_{\mathcal{A}(x,t)}(k)}e^{i\alpha_{j,j+1}n_\mathcal{A}(x,t,k)},
\end{equation}
and, importantly, the restriction $n_\mathcal{A}$ of the charge operator $N_A$ (cf. Eq.~\eqref{eq:chmom_gamma}) to the block $\mathcal{A}$ has to be written in the same basis 
as $G_\mathcal{A}$, i.e., in  the basis of the fermionic modes $d_j$ (cf.~\eqref{eq:gen-quasipart_1}). For the models and quenches discussed here, 
$n_\mathcal{A}$ remains diagonal as $n_\mathcal{A}=\mathrm{diag}(-1,\dots,-1,1,\dots,1)$, although this is not true in 
general.
We now show that~\eqref{eq:chmom_quasipart} yields 
Eqs.~\eqref{eq:chmom_general} with~\eqref{eq:nodiss_t0} and~\eqref{eq:nodiss_tinf} 
for the $XX$ chain, and it correctly reproduces known results for the dynamics 
of the asymmetry after quenches in the $XX$ chain~\cite{ares2023lack,murciano2023entanglement}.

\subsubsection{First example: quench from the tilted N\'eel state in the $XX$ chain.}\label{sec:quench-XX}

As a first example of the correctness of~\eqref{eq:chmom_quasipart} here we discuss 
the quench from the tilted N\'eel state in Eq.~\eqref{eq:cat} with the $XX$ 
Hamiltonian~\eqref{eq:XX_diag}. To apply the multiplet framework,  we first define 
the fermionic operators associated with the different species $d_1(k):=c_k$, $d_2(k):=c_{\pi-k}$, $d_3(k):=c_{\pi+k}$, $d_4(k):=c_{2\pi-k}$, with $c_k$ the Fourier transform of the original fermions (see Section~\ref{sec:models}). 
The $XX$ Hamiltonian~\eqref{eq:XX_diag} can be rewritten as 
\begin{equation}
    \label{eq:XX_multiplets}
    H=\int_0^{\frac{\pi}{2}}\frac{dk}{2\pi} \sum_{j=1}^4 \varepsilon_j(k) d_j^\dagger(k)d_j(k),
\end{equation}
with $\varepsilon_1(k)=-\cos k$, $\varepsilon_2(k)=-\cos(\pi-k)$, $\varepsilon_3(k)=-\cos(\pi+k)$, $\varepsilon_4(k)=-\cos(2\pi-k)$. The number of quasiparticles forming the entangled multiplet is dictated by the structure of the initial state. For the tilted N\'eel state the quadruplet structure arises from the 
fact that the expectation values of both the correlators $c^\dagger_x c_y$ and $c_x c_y$ are nonzero, and because of the 
two-site periodicity of the initial state. According to our discussion in section~\ref{sec:mult-quasi}, the velocities of the four species are $v_1(k)=v_2(k)=\sin k$ and $v_3(k)=v_4(k)=-\sin k$. Notice that $v_2(k)=-d\varepsilon_2(k)/dk$ and $v_4(k)=-d\varepsilon_4(k)/dk$ because 
for the species with $j=2$, one has $k'=\pi-k$ and for $j=4$, one has $k'=2\pi-k$. The two-point fermionic correlation functions calculated over the tilted N\'eel state are block-diagonal in terms of $d_j(k)$, which is the reason for introducing them. Indeed, the only nonzero correlators in the $k$-space are  
\begin{align}
    \label{eq:cdagc_corr}
        & \langle\Psi(0)| c_k^\dagger c_k |\Psi(0)\rangle=\frac{1}{2} (1+g_{12}(2k,\theta)),\\
        \label{eq:cdagc_corr-1}
        & \langle\Psi(0)| c_k^\dagger c_{\pi+k} |\Psi(0)\rangle=-\frac{1}{2} g_{11}(2k,\theta),
\end{align}
and
\begin{align}
    \label{eq:cc_corr}
        & \langle \Psi(0)|c_k c_{2\pi-k}|\Psi(0)\rangle=-\frac{i}{2}f_{12}(2k,\theta),\\
        \label{eq:cc_corr-1}
        & \langle \Psi(0)|c_k c_{\pi-k}|\Psi(0)\rangle=\frac{i}{2}f_{11}(2k,\theta).
\end{align}
The functions $f_{ij}$ and $g_{ij}$ appearing in the correlators in~\eqref{eq:cdagc_corr},~\eqref{eq:cdagc_corr-1} and~\eqref{eq:cc_corr},~\eqref{eq:cc_corr-1} are the same as in Eq.~\eqref{eq:g11}. Notice the rescaling $k\to 2k$ as compared with~\eqref{eq:g11}, which reflects that the Brillouin zone is halved for the 
tilted N\'eel state. Again, only $c_k,c_{\pi\pm k},c_{2\pi-k}$ appear 
in~\eqref{eq:cdagc_corr}~\eqref{eq:cdagc_corr-1}, which makes apparent 
the reason for our definitions of the different species. 
We can now write down $G(k)$  (cf.~\eqref{eq:gen-quasipart_3}) in block diagonal form.  The 
blocks $C(k)$ and $F(k)$  in the basis $d_1(k),d_2(k),d_3(k),d_4(k)$ read 
\begin{equation}
    \label{eq:C(k)}
    C(k)=\begin{pmatrix}
         g_{12}(2k,\theta)&0&-g_{11}(2k,\theta)&0\\
        0&- g_{12}(2k,\theta)&0&-g_{11}(2k,\theta)\\
        - g_{11}(2k,\theta)&0&-g_{12}(2k,\theta)&0\\
        0&- g_{11}(2k,\theta)&0&g_{12}(2k,\theta)\\
    \end{pmatrix},
\end{equation}
and
\begin{equation}
    \label{eq:F(k)}
    F(k)=\begin{pmatrix}
        0&if_{11}(2k,\theta)&0&-if_{12}(2k,\theta)\\
        -if_{11}(2k,\theta)&0&-if_{12}(2k,\theta)&0\\
        0&i f_{12}(2k,\theta)&0&if_{11}(2k,\theta)\\
        if_{12}(2k,\theta)&0&-if_{11}(2k,\theta)&0
    \end{pmatrix}.
\end{equation}
We stress that $C(k)$ and $F(k)$ are different from $\mathcal{C}(k)$ and $\mathcal{F}(k)$ in 
Eqs.~\eqref{eq:ckt} and~\eqref{eq:fkt}, as can be seen from the definition of $G(k)$ based on Eq.~\eqref{eq:gen-quasipart_2} (see~\cite{ares2023lack}). Clearly, $C(k)$ and $F(k)$ are not diagonal, signaling nontrivial correlation and 
entanglement between the different species of quasiparticles. 

To obtain the quasiparticle picture for the charged moments one has to determine the kinematics of the 
quasiparticles. Since we observed that $v_1=v_2$ and $v_3=v_4$, the contribution of the quadruplet to the charged moment can originate from the configurations in which two quasiparticles forming the multiplet are in $A$ (either $1,2$ or $3,4$). The other possibility is that all four quasiparticles are in $A$ (indeed, the configuration with no quasiparticles in $A$ immediately yields zero). We anticipate that this configuration gives a non-trivial contribution, differently from the quasiparticle picture for the 
entanglement entropy, for which only quadruplets that are shared between the subsystem and the rest contribute to the 
entanglement between them.

The fact that the quadruplets $d_1,d_2,d_3,d_4$ are formed by two pairs of quasiparticles with the same velocity, prevents the symmetry restoration by contributing non-trivially to the charged moments and to the integrals in Eq.~\eqref{eq:mpemba_neel} at any time $t$.  Physically, the fact that $v_1=v_2$ and $v_3=v_4$ implies that 
even at infinite time there are always pairs of ``correlated'' fermions in $A$, which prevent the Cooper pairs 
correlator from vanishing (see also \cite{ares2023lack} for an explanation of the lack of restoration of the $U(1)$ symmetry in terms of a non-Abelian generalised Gibbs ensemble).  
Let us now observe that the total number of multiplets with the pair $1-2$ or $3-4$ of quasiparticles in $A$ are $\min(\ell, 2|v_1(k)|t)$ and $\min(\ell,2|v_3(k)|)$, respectively. On the other hand, the number of multiplets with  all the four quasiparticles  inside $A$ is $\max\left(\ell-2|v_1(k)|t,0\right)$.

Putting everything together and using~\eqref{eq:chmom_quasipart}, we find the predicted leading order behaviour of the charged moments as
\begin{equation}
    \label{eq:chmom_quasipart_neel}
    \ln Z_n (\boldsymbol{\alpha},t)= \int_0^{\frac{\pi}{2}} \frac{dk}{2\pi} \left[\max\left(\ell-2|v_1(k)|t,0\right) z_n^{(4)}(k) + \min(\ell, 2|v_1(k)|t)\; (z_n^{\{1,2\}}(k)+z_n^{\{3,4\}}(k))\right],
\end{equation}
where $z_n^{\scriptscriptstyle(4)}(k)$ is the multiplet contribution~\eqref{eq:chmom_quasipart} for all the four quasiparticles inside the subsystem, while $z_n^{\scriptscriptstyle\{1,2\}}(k)$, $z_n^{\scriptscriptstyle\{3,4\}}(k)$ are the contributions for the case with  quasiparticles $1,2$ and $3,4$ in $A$. 

We now explicitly show that Eq.~\eqref{eq:chmom_quasipart_neel} coincides with the quasiparticle picture~\eqref{eq:chmom_general} for the charged moments reported in Sec.~\ref{subsec:nodiss}.
Let us give a closer look at Eq.~\eqref{eq:chmom_quasipart_neel}. First, 
$z^{\scriptscriptstyle(4)}$ is given as (cf.~\eqref{eq:chmom_quasipart}) 
\begin{equation}
    \label{eq:quasipart_neel_z4_1}
    z_n^{4}(k)=\frac{1}{2} \ln\left[\det\left(\left(\frac{\mathds{1}-G(k)}{2}\right)^n\left(\mathds{1}+\prod_{j=1}^n w_j(x,t,k)\right)\right)\right], 
\end{equation}
where $G(k)$ is defined in~\eqref{eq:gen-quasipart_3}, and $C(k)$ and $F(k)$ are given in~\eqref{eq:C(k)} and~\eqref{eq:F(k)}. 
We can reorder rows and columns of the $8\times 8$ matrix $G(k)$ into a block-diagonal form with $4\times 4$ blocks, such that Eq.~\eqref{eq:quasipart_neel_z4_1} becomes  
\begin{multline}
    \label{eq:quasipart_neel_z4_2}
        z_n^{(4)}(k)=\frac{1}{2} \ln\left[\det\left(\left(\frac{\mathds{1}-G^{(1)}(k)}{2}\right)^n\left(\mathds{1}+\prod_{j=1}^n w_j^{(1)}(x,t,k)\right)\right)\right]+\\
        +\frac{1}{2} \ln\left[\det\left(\left(\frac{\mathds{1}-G^{(2)}(k)}{2}\right)^n\left(\mathds{1}+\prod_{j=1}^n w_j^{(2)}(x,t,k)\right)\right)\right],
\end{multline}
with
\begin{equation}
    \label{eq:quasipart_neel_z4_3}
    G^{(1)}(k):=\begin{pmatrix}
        g_{12}(2k,\theta)&-g_{11}(2k,\theta)&if_{11}(2k,\theta)&-if_{12}(2k,\theta)\\
        -g_{11}(2k,\theta)&-g_{12}(2k,\theta)&if_{12}(2k,\theta)&if_{11}(2k,\theta)\\
        -if_{11}(2k,\theta)&-if_{12}(2k,\theta)&g_{12}(2k,\theta)&g_{11}(2k,\theta)\\
        if_{12}(2k,\theta)&-if_{11}(2k,\theta)&g_{11}(2k,\theta)&-g_{12}(2k,\theta)
    \end{pmatrix}
\end{equation}
and $G^{(2)}(k):=G^{(1)}(\pi-k)$. Here $w_j^{(1)}$,$w_j^{(2)}$ are defined in~\eqref{eq:wj}, where now $n_\mathcal{A}$ is a $4\times4$ matrix. One can check that $G^{(1)}(k)$ is isospectral to $\gcal_0(2k)$ (cf.~\eqref{eq:G_pm0}) and $\left(\mathds{1}+G^{(1)}(k)\right)n_\mathcal{A}$ to $\left(\mathds{1}+\gcal_0(2k)\right)n_A$ (cf.~\eqref{eq:G_pm0}). This allows us to  obtain (see the proof of Eq.~\eqref{eq:nodiss_t0} in Appendix~\ref{app:proof_nodiss})
\begin{equation}
    \label{eq:quasipart_neel_z4_4}
        z_n^{(4)}(k)=\frac{1}{2} \ln\left(\det \mathcal{Z}_{n,0}^{(\boldsymbol{\alpha})}(2k)\right)+ \frac{1}{2} \ln\left(\det\mathcal{Z}_{n,0}^{(\boldsymbol{\alpha})}(2\pi-2k)\right).
\end{equation}
The contribution $z_n^{\scriptscriptstyle\{1,2\}}$ corresponding to two of the quasiparticles 
forming the quadruplet being in $A$ is, instead, 
\begin{equation}
    \label{eq:quasipart_neel_z2_1}
    z_n^{\{1,2\}}(k)=\frac{1}{2} \ln\left[\det\left(\left(\frac{\mathds{1}-G_{\{1,2\}}(k)}{2}\right)^n\left(\mathds{1}+\prod_{j=1}^n w_j^{\{1,2\}}(x,t,k)\right)\right)\right],
\end{equation}
with the restricted two-point function $G_{\{1,2\}}(k)$ reading
\begin{equation}
    \label{eq:quasipart_neel_z2_2}
    G_{\{1,2\}}(k)=\begin{pmatrix}
        g_{12}(2k,\theta)&0&0&if_{11}(2k,\theta)\\
        0&-g_{12}(2k,\theta)&-if_{11}(2k,\theta)&0\\
        &if_{11}(2k,\theta)&-g_{12}(2k,\theta)&0\\
        -if_{11}(2k,\theta)&0&0&g_{12}(2k,\theta)
    \end{pmatrix}.
\end{equation}
The matrix $G_{\{1,2\}}$ is obtained from~\eqref{eq:gen-quasipart_3} by selecting rows and columns 
corresponding to the quasiparticles in $A$. 
The contribution of the configuration with quasiparticles $3,4$ in $A$ reads $z_n^{\scriptscriptstyle\{3,4\}}(k)=z_n^{\scriptscriptstyle\{1,2\}}(\pi-k)$, since $G_{\{3,4\}}(k)=G_{\{1,2\}}(\pi-k)$. As for $z_n^{\scriptscriptstyle(4)}$, it is easy to show that $G_{\{1,2\}}(k)$ is isospectral to $\gcal_\infty(2k)$ (cf.~\eqref{eq:G_pm0}) and that, in the eigenbasis of $G_{\{1,2\}}(k)$, we can rewrite $n_\mathcal{A}=\sigma_x \otimes \mathds{1}$. Thus, by using the same steps in the proof of Eq.~\eqref{eq:nodiss_tinf} in Appendix~\ref{app:proof_nodiss}, we obtain 
\begin{equation}
    \label{eq:quasipart_neel_z2_3}
    z_n^{\{1,2\}}(k)+z_n^{\{3,4\}}(k)=\frac{1}{2} \ln\left[\det \mathcal{Z}_{n,\infty}^{(\boldsymbol{\alpha})}(2k)\right]+ \frac{1}{2} \ln\left[\det\mathcal{Z}_{n,\infty}^{(\boldsymbol{\alpha})}(2\pi-2k)\right].
\end{equation}
If we finally substitute  Eqs.~\eqref{eq:quasipart_neel_z4_4} and~\eqref{eq:quasipart_neel_z2_3} in Eq.~\eqref{eq:chmom_quasipart_neel}, and we change the integration variable as $k'=2k$, we recover Eq.~\eqref{eq:chmom_general}, proving the equivalence of~\eqref{eq:chmom_quasipart_neel} and~\eqref{eq:chmom_general}. 

\subsubsection{Second example: Quench from XY spin chain}
\label{sec:quench-XY}

Here we consider the quench in the $XX$ chain starting from the ground state of the $XY$ chain. 
We employ~\eqref{eq:chmom_general} to recover  the quasiparticle picture for the dynamics of the charged moments, which was derived in~\cite{murciano2023entanglement}. 
 To employ our generalized quasiparticle picture, we first observe that the quench produces only pairs of entangled quasiparticles. Let us define $d_1(k):=c_k$ and $d_2(k):=c_{2\pi-k}$, 
 with $k\in[0,\pi]$.  We now have $\varepsilon_1(k)=\varepsilon_{XX}(k)$ and $\varepsilon_2(k)=\varepsilon_{XX}(2\pi-k)$ (cf.~\eqref{eq:XY_diag}). This yields opposite velocities $v_1(k)=-v_2(k)=\varepsilon'(k)$. The two-point fermionic correlation function on the initial state, 
 i.e., the ground state $|\Omega\rangle$ of the $XY$ chain, is block diagonal with respect to these two species of quasiparticles. Specifically, the only two nonzero correlators in the quasimomentum space are
\begin{equation}
    \label{XY_corr}
    \langle\Omega|c_k^\dagger c_k|\Omega\rangle=\sin^2(\Delta_k/2) \qquad \text{and} \qquad \langle\Omega|c_k c_{2\pi-k}|\Omega\rangle=-i/2 \sin(\Delta_k).
\end{equation}
Thus, the correlation matrix $G(k)$ (cf.~\eqref{eq:gen-quasipart_3}) reads
\begin{equation}
\label{eq:G-supp}
G(k)=\begin{pmatrix}
C(k) & F^\dagger(k)\\
F(k) & -C^T(k)
\end{pmatrix},
\end{equation}
with
\begin{equation}
\label{eq:matrix-supp}
C(k)=\begin{pmatrix}
-\cos(\Delta_k) & 0\\
0 & -\cos(\Delta_k)
\end{pmatrix}, \qquad
F(k)=\begin{pmatrix}
0 & -i\sin(\Delta_k)\\
i\sin(\Delta_k) &0
\end{pmatrix},
\end{equation}
where the row and column indices of the matrices $G(k)$ and $F(k)$ identify the two species. 
The leading order of the charged moments in the hydrodynamic limit 
for an interval $A$ of size $\ell$ was obtained in Ref.~\cite{murciano2023entanglement} as
\begin{multline}
    \label{eq:XY_chmom_analytic}
   \ln Z_n (\boldsymbol{\alpha},t)=\\  \int_{0}^{2\pi}\frac{dk}{2\pi}\left[ \frac{1}{2} \max\left(\ell-2|v_1(k)|t,0\right)\ln \prod_{j=1}^n f_k(\alpha_{j,j+1})+\min (2|v_1(k)|t,\ell)\ln h_n(n(k))\right],
\end{multline}
where 
\begin{align}
\label{eq:fk}
& f_k(\alpha):=\cos \alpha + i \cos \Delta_k \sin \alpha\\
\label{eq:nk}
& n(k):=\frac{1}{2}(1-\cos\Delta_k).
\end{align}
We now rederive~\eqref{eq:XY_chmom_analytic} by using the general formula~\eqref{eq:chmom_quasipart}. 
Since the quench produces only entangled pairs, one has  configurations with 
no quasiparticles in $A$, or one or two quasiparticles in $A$. If no quasiparticles are present in $A$, the contribution to the charged moments is trivially zero.  If one or two quasiparticles are in $A$ there are nonzero contributions that we denote as $z_n^{\{1\}}(k,\boldsymbol{\alpha})$ and  $z_n^{\{2\}}(k,\boldsymbol{\alpha})$, respectively. The number of configurations corresponding to one or both quasiparticles in $A$ is $\min(\ell,2|v_1(k)|t)$  and $\max(\ell-2|v_1(k)|t)$, respectively. Notice that 
they are the same as in~\eqref{eq:chmom_quasipart_neel}. In conclusion, we obtain 
\begin{multline}
    \label{eq:chmom_quasipart_XY}
    \ln Z_n (\boldsymbol{\alpha},t)= \int_0^{\pi} \frac{dk}{2\pi} \left[\max\left(\ell-2|v_1(k)|t,0\right) z_n^{\{1,2\}}(k,\boldsymbol{\alpha}) +\right.\\
    \left.+\min(\ell, 2|v_1(k)|t)\; \left(z_n^{\{1\}}(k,\boldsymbol{\alpha})+z_n^{\{2\}}(k,\boldsymbol{\alpha})\right)\right].
\end{multline}
Let us now consider the matrix $G_{\{1\}}=G_{\{2\}}$, which is obtained by restricting the 
row and column indices to $1$ (or $2$) in~\eqref{eq:matrix-supp} and using~\eqref{eq:G-supp}. 
It reads as 
\begin{equation}
    \label{eq:C1_XY}
    G_{\{1\}}(k)=G_{\{2\}}(k)=\begin{pmatrix}
         -\cos \Delta_k&0\\
        0&\cos \Delta_k
    \end{pmatrix}. 
\end{equation}
Since $G_{\{1\}}$ and $G_{\{2\}}$ commute with the  matrix 
$n_\mathcal{A}=-\sigma_z$, the one-particle contributions $z_n^{\scriptscriptstyle\{2\}}$ and 
$z_n^{\scriptscriptstyle\{2\}}$ are independent from $\boldsymbol{\alpha}$.  
It is straightforward  to obtain that 
\begin{equation}
    \label{eq:z1_XY}
    z_n^{\{1\}}(k,\boldsymbol{\alpha})=z_n^{\{2\}}(k,\boldsymbol{\alpha})=\ln h_n(n(k)), 
\end{equation}
where $h_n(x)$ is defined in~\eqref{eq:hn} and $n(k)$ in~\eqref{eq:nk}. 
On the other hand, the contribution $z_n^{\scriptscriptstyle\{1,2\}}$ due to both members of the pair  
being in $A$ can be derived following the same steps to obtain Eq.~\eqref{eq:nodiss_t0} (see Appendix~\ref{app:proof_nodiss}). 
The result reads as 
\begin{equation}
    \label{eq:z2_XY}
    z_n^{\{1,2\}}(k,\boldsymbol{\alpha})=\ln \prod_{j=1}^n f_k(\alpha_{j,j+1}),
\end{equation}
where $f_k$ is defined in~\eqref{eq:fk}. 
We report the derivation of~\eqref{eq:z2_XY} in Appendix~\ref{app:XY_proof}. Finally, after substituting   Eqs.~\eqref{eq:z1_XY} and~\eqref{eq:z2_XY} in Eq.~\eqref{eq:chmom_quasipart_XY}, and using that the integrand in~\eqref{eq:XY_chmom_analytic} is invariant under $k\to2\pi-k$, we obtain formula~\eqref{eq:XY_chmom_analytic}. 

The results derived above allow us to draw some general conclusions on the dynamics of the charged moments and the asymmetry after generic quenches producing entangled pairs of quasiparticles. 
As anticipated, in contrast with the von Neumann and R\'enyi entropies, when an entangled pair is shared between the subsystem $A$ and its complement $\overline{A}$, typically it does \textit{not} contribute to the entanglement asymmetry $\Delta S_A^{(n)}$.  For the quench from the ground state of the $XY$ chain this is clear because  the asymmetry depends on the ratio $\left(Z_n(\boldsymbol{\alpha})/Z_n(0)\right)$ (cf.~\eqref{eq:asym_from_chmom}). Now, since~\eqref{eq:z1_XY} does not depend on $\boldsymbol{\alpha}$, Eq.~\eqref{eq:z1_XY} gives no contribution to the asymmetry. 
Interestingly, this happens for \textit{any} free-fermion quench producing entangled pairs travelling with opposite velocities and such that the restricted $2\times2$ correlation matrix $G_{\{j\}}$ commutes with the charge operator $n_\mathcal{A}$. Since both $G_{\{j\}}$ and $n_\mathcal{A}$ are $2\times 2$ matrices, it can be easily shown that the commutation condition is equivalent to $n_\mathcal{A}$ being diagonal; 
 it ensures that $z_n^{\{j\}}(k,\boldsymbol{\alpha})=\ln h_n (n_j(k))$, where $n_j(k)$ is the density of quasiparticles of type $j$ produced after the quench.  This implies that the only  nontrivial contribution to the entanglement asymmetry originates  from pairs fully contained in the subsystem. 
 At infinite time there are no pairs with both particles contained in any finite subsystem $A$, 
implying that for quenches producing entangled pairs, the symmetry is \textit{always} asymptotically restored, provided that $[G_{\{j\}},n_\mathcal{A}]=0$. Notice that this reasoning can be straightforwardly generalized to quenches producing larger multiplets, if all the species of the multiplet propagate with different velocities: also in these cases the symmetry is asymptotically restored.

\section{Dissipation  and the quantum Mpemba effect}\label{sec:results}

Here we discuss the effect of dissipation on the symmetry restoration and the quantum Mpemba 
effect. We focus on the quench from the tilted N\'eel state (cf.~\eqref{eq:tiltedNéel}) in the $XX$ chain in the presence of 
gain and loss dissipation. Crucially, in the absence of dissipation,  the $U(1)$ symmetry of the $XX$ chain is not 
restored. This can be interpreted as the consequence of non-Abelian conservation laws~\cite{fagotti2014on} of the Hamiltonian \cite{ares2023lack}. Here we show that in the presence of dissipation 
the $U(1)$ symmetry of the $XX$ chain is restored. Concomitantly, we show that quantum Mpemba effect 
is robust against dissipation.

Let us first remark that by using  Eq.~\eqref{eq:chmom_general_diss} for the charged moments in 
the weakly-dissipative hydrodynamic limit 
and the definitions~\eqref{eq:FT} and~\eqref{eq:def_renyi}, it is straightforward to compute, 
at least numerically, the entanglement asymmetry for generic dissipation.

In the following, we first analytically derive the leading order of the entanglement asymmetry 
$\Delta S_A^{(n)}$. Specifically, for balanced gain and losses we provide the formula for 
$\Delta S_A^{(n)}$ in the limit $t\to\infty$ for any $n\neq 1$. For $\gamma^+\ne \gamma^-$ we provide the 
result for $\Delta S_A^{(2)}$ in the limit $t\to\infty$. Moreover, by comparing the short- and 
long-time behaviour of the asymmetry, we obtain a condition for the Mpemba effect in the presence of 
gain/loss dissipation. In Section~\ref{sec:quasi-mpemba} we compare this condition with the case without 
dissipation. 

When $t=0$, i.e., before the quench, the entanglement asymmetry is the same as in the 
case without dissipation, which was already derived in~\cite{ares2023lack} for large 
subsystem size $\ell$. The result reads as~\cite{ares2023lack} 
\begin{equation}
    \label{eq:asym_leading_t0}
    \Delta S^{(n)}_A(\theta,t=0)=\frac{1}{2} \ln \ell + \frac{1}{2}\ln\left[\frac{\pi n^{1/(n-1)}}{4} \int_0^{2\pi}\frac{dk}{2\pi}\left(f_{11}(k,\theta)^2+f_{12}(k,\theta)^2\right)\right]+ O(\ell^{1-n}),
\end{equation}
with $f_{ij}$ defined in~\eqref{eq:g11}. 
To determine the behaviour of the asymmetry in the limit $t\to \infty$, 
let us start from the ratio of the charged moments in the hydrodynamic limit
\begin{multline}
    \label{eq:asym_general}
       \frac{Z_n(\boldsymbol{\alpha},t)}{Z_n(0,t)}=\\\exp \left\{ \frac{\ell}{4} \int_0^{2\pi} \frac{dk}{2\pi} \left[\ln\frac{\det\mathcal{Z'}_{n,0}^{(\mathbf{\alpha})}}{\det\mathcal{Z'}_{n,0}^{(\mathbf{0})}} \max\left(1-2|v_1(k)|t/\ell,0\right)+ \min(1,2|v(k)| t/\ell)\ln \frac{\det\mathcal{Z'}_{n,\infty}^{(\boldsymbol{\alpha})}}{\det\mathcal{Z,}_{n,\infty}^{(\mathbf{0})}}\right] \right\}.
\end{multline}
As it is clear from the definitions~\eqref{eq:asy-first}~\eqref{eq:rhoa-1}~\eqref{eq:FT}~\eqref{eq:Znalpha}, Eq.~\eqref{eq:asym_general} allows one to obtain the 
entanglement asymmetry. In the large-time limit $t/\ell \to \infty$, only the term in~\eqref{eq:asym_general} multiplying $\min(1,2|v(k)| t/\ell)$ survives. Thus,  by plugging the result in Eq.~\eqref{eq:FT}, we find that the entanglement asymmetry is 
\begin{equation}
    \label{eq:asym_general_2}
       \Delta S^{(n)}_A(\theta,t \to \infty) \simeq \frac{1}{1-n}\ln\left\{\int_{-\pi}^\pi \frac{d\alpha_1\dots d\alpha_n}{(2\pi)^n}\exp \left[ \frac{\ell}{4} \int_0^{2\pi} \frac{dk}{2\pi} \ln \frac{\det\mathcal{Z}_{n,\infty}^{(\boldsymbol{\alpha})}}{\det\mathcal{Z}_{n,\infty}^{(\mathbf{0})}}\right]\right\}.
\end{equation}
To proceed, let us focus on the case with  $n=2$.  If we plug Eq.~\eqref{eq:z2_generic_tinf} in Eq.~\eqref{eq:asym_general_2} and we expand at the leading order in the small parameter $\lambda(t)$, we find 
\begin{equation}
    \label{eq:asym_n2_1}
    \Delta S^{(2)}_A(\theta,t \to \infty) \simeq -\ln \left\{\int_{-\pi}^{\pi}\frac{d\alpha}{2\pi} \exp \left[-\lambda(t)^2 \frac{2\ell}{(1+s^2)^2} \left(\int_0^{2\pi} \frac{dk}{2\pi} f_{11}(k,\theta)^2\right) \sin^2(\alpha)\right]\right\},
\end{equation}
where $\alpha=\alpha_1-\alpha_2$, and $s=(\gamma^+-\gamma^-)/(\gamma^++\gamma^-)$. If we define 
\begin{equation}
    \label{eq:A}
    A_2:=\lambda(t)^2 \frac{2\ell}{(1+s^2)^2} \int_0^{2\pi} \frac{dk}{2\pi} f_{11}(k,\theta)^2,
\end{equation}
the remaining integration in $\alpha$ yields
\begin{equation}
    \label{eq:asym_n2_2}
    \Delta S^{(2)}_A(\theta,t \to \infty) \simeq -\ln\left[ e^{-\frac{A_2}{2}} I_0\left(\frac{A_2}{2}\right)\right]\simeq \frac{A_2}{2},
\end{equation}
where $I_0$ is the modified Bessel function. Finally, we obtain 
\begin{equation}
    \label{eq:asym_n2_3}
    \Delta S^{(2)}_A(\theta,t \to \infty) \simeq \lambda(t)^2  \frac{(\gamma_++\gamma_-)^4}{4(\gamma_+^2+\gamma_-^2)^2} \ell \int_0^{2\pi} \frac{dk}{2\pi} f_{11}(k,\theta)^2.
\end{equation}
Let us now consider the case $\gamma_+=\gamma_-$. We can evaluate the leading order behaviour of the R\'enyi entanglement asymmetry for any $n \neq 1$, by replacing Eq.~\eqref{eq:zn_gpgm_tinf} into Eq.~\eqref{eq:asym_general_2} and expanding at the leading order in $\lambda(t)$, such that we obtain 
\begin{equation}
    \label{eq:asym_gpgm_1}
    \begin{gathered}
    \Delta S^{(n)}_A(\theta,t \to \infty) \simeq \frac{1}{1-n}\ln\left\{\int_{-\pi}^\pi \frac{d\alpha_1\dots d\alpha_n}{(2\pi)^n} \exp \left[ -\lambda(t)^2 2 \ell  \right. \right.\\
    \left.\left.\times \left(\int_0^{2\pi} \frac{dk}{2\pi} f_{11}(k,\theta)^2\right) \sum_{1\leq p_1<p_2\leq n} \sin^2(\alpha_{p_1}-\alpha_{p_2}) \right]\right\}
    \end{gathered}
\end{equation}
We stress that the result above holds for $n\neq 1$ because we are not able to analytically continue in $n$ Eqs. \eqref{eq:zn_gpgm_tinf} and \eqref{eq:angles_tinf}. Thus, we cannot prove that the limits $n\to 1$ and $t\to \infty$ commute (see for example \cite{murciano2023entanglement}). 
Using that $\lambda(t)=e^{-2\gamma_+t}\to 0$ in the large time limit, we can expand the exponential in the integral above.
The resulting $n$-fold integral in $\boldsymbol{\alpha}$ can be computed by observing that $\int dx dy \sin^2(x-y)=2\pi^2$ and, since the sum over $p_j$'s involves $\binom{n}{2}$ terms, we get 
\begin{equation}
    \int_{-\pi}^\pi \frac{d\alpha_1\dots d\alpha_n}{(2\pi)^n}\sum_{1\leq p_1<p_2\leq n} \sin^2(\alpha_{p_1}-\alpha_{p_2}) =\frac{1}{2}\binom{n}{2}.
\end{equation}
This allows us rewrite  Eq.~\eqref{eq:asym_gpgm_1}  as
\begin{equation}
    \label{eq:asym_gpgm_S}
    \Delta S^{(n)}_A(\theta,t \to \infty) \simeq \frac{1}{2(n-1)}\binom{n}{2}A_2.
\end{equation}
As we anticipated, our goal is to determine the condition under which the Mpemba effect occurs.  
Given two initial tilted N\'eel states with tilting angles $\theta_1>\theta_2$, in formulas, this 
condition reads 
\begin{equation}
    \label{eq:mpemba}
    \left\{\begin{array}{c}
        \Delta S^{(n)}_A(\theta_1,0)>\Delta S_A^{(n)}(\theta_2,0),\\
        \Delta S^{(n)}_A(\theta_1,t\to \infty)<\Delta S_A^{(n)}(\theta_2,t\to \infty).
        \end{array}
    \right.
\end{equation}
For all the cases in which we were able to explicitly compute the entanglement asymmetry, by using  Eqs.~\eqref{eq:asym_leading_t0},~\eqref{eq:asym_n2_3} and~\eqref{eq:asym_gpgm_S}, the 
condition~\eqref{eq:mpemba} becomes 
\begin{equation}
    \label{eq:mpemba_diss}
    \left\{
    \begin{gathered}
        \int_0^{2\pi}\frac{dk}{2\pi}\left(f_{11}(k,\theta_1)^2+f_{12}(k,\theta_1)^2\right)>\int_0^{2\pi}\frac{dk}{2\pi}\left(f_{11}(k,\theta_2)^2+f_{12}(k,\theta_2)^2\right),\\
        \int_0^{2\pi} \frac{dk}{2\pi} f_{11}(k,\theta_1)^2<\int_0^{2\pi} \frac{dk}{2\pi} f_{11}(k,\theta_2)^2.
    \end{gathered}
    \right. 
\end{equation}
We notice that, for $\theta \in [0,\pi/2]$, the first condition is always satisfied as far as $\theta_1>\theta_2$. One expects the criterion~\eqref{eq:mpemba_diss} to hold also for generic gain/loss processes, although we were not able to prove it. 
In the following, we compare the condition~\eqref{eq:mpemba_diss} with the one obtained in other (non-dissipative) free-fermion quench protocols~\cite{murciano2023entanglement}. 

\subsection{Multiplet picture and onset of Mpemba: Comparison with the non-dissipative case}
\label{sec:quasi-mpemba}

Here we show that the generalized quasiparticle picture for the charged moments introduced in 
section~\ref{sec:chmom_quasipart} allows to understand the condition~\eqref{eq:mpemba_diss} giving rise to the 
Mpemba effect both in the non-dissipative case, as well as in the presence of dissipation.

We can now revisit the condition~\eqref{eq:mpemba_diss} under which the Mpemba effect occurs. We start discussing the 
condition for Mpemba in the absence of dissipation. We first consider the quench from the ground state of the $XY$ spin chain in the $XX$ chain. Then we focus on the quench in the $XX$ chain starting from the  tilted N\'eel state.
For the latter (cf.~\eqref{eq:cat}) in the absence of dissipation 
the symmetry is not asymptotically restored (unless $\theta=\pi/2$~\cite{ares2023lack}). Therefore, quantum Mpemba effect 
does not occur. 
Nevertheless, even though the symmetry is not restored, a crossing between the curves  for the asymmetry corresponding to different 
tilting angles can occur. We can still write the condition~\eqref{eq:mpemba} which ensures that there is a crossing in time between $\Delta S_A^{(n)}(\theta_1,t)$ and $\Delta S_A^{(n)}(\theta_2,t)$. 
The main goal here is to show that the condition for Mpemba to happen in the presence of dissipation is similar to 
the crossing condition in the absence of symmetry restoration. Specifically, in these conditions all the quasimomenta contribute, in contrast with the condition for Mpemba found in~\cite{rylands2023microscopic,murciano2023entanglement}.

Let us first discuss the condition for the standard Mpemba effect,  focussing on the quench from the ground state of the $XY$ chain (see section~\ref{sec:models} and~\ref{sec:chmom_quasipart}). 
The post-quench dynamics is driven by the $XX$ Hamiltonian. The condition to observe the Mpemba effect 
discussed in Ref.~\cite{murciano2023entanglement} can be interpreted in terms of multiplets. 
For the quench from the ground state of the $XY$ chain only pairs $(k,2\pi-k)$ are produced. 
First, at time $t=0$, the leading order contribution to the asymmetry is~\cite{murciano2023entanglement} 
\begin{equation}
    \label{eq:asym_XY_t0}
    \Delta S^{(n)}(\gamma,h,t=0)=\frac{1}{2} \ln \ell + \frac{1}{2}\ln\left(\frac{\pi n^{1/(n-1)}}{4} \int_0^{2\pi} \frac{dk}{2\pi} \sin^2 \Delta_k(\gamma,h)\right)+ O(\ell^{1-n}),
\end{equation}
where $\Delta_k(\gamma,h)$ is the Bogoliubov angle in Eq.~\eqref{eq:Bogoliubov_angle}, while at long time it becomes
\begin{equation}
    \label{eq:asym_XY_tinf}
    \Delta S^{(n)}(\gamma,h,t\to \infty)\simeq \frac{n\ell}{1-n}\int_0^{2\pi} \frac{dk}{16\pi} \left[1-\min\left(\frac{2|v(k)|t}{\ell},1\right)\right] \sin^2 \Delta_k (\gamma,h).
\end{equation}
Thus, the quantum Mpemba effect occurs if there exists a time, $t_I$, such that~\cite{murciano2023entanglement}
\begin{equation}
    \label{eq:mpemba_XY}
    \begin{cases}
        \displaystyle \int_{0}^{2\pi}\frac{dk}{2\pi}\sin^2\Delta_k(\gamma_1,h_1)>\int_{0}^{2\pi}\frac{dk}{2\pi}\sin^2\Delta_k(\gamma_2,h_2),\\
    \qquad  \displaystyle   \int_{-k*}^{k*} \frac{dk}{2\pi} \Upsilon_k(\gamma_1,h_1)<\int_{-k*}^{k*} \frac{dk}{2\pi} \Upsilon_k(\gamma_2,h_2),\qquad \text{for } t>t_I
    \end{cases}
\end{equation}
where $\Upsilon_k(\gamma,h):=\sin^2\Delta_k(\gamma,h)+\sin^2\Delta_{k+\pi}(\gamma,h)$ and $k^*:=\arcsin[{t/(2\ell)}]$. 
The first inequality does not depend on whether the dynamics is unitary or dissipative. On the other hand, the second condition in~\eqref{eq:mpemba_XY} involves only the modes 
in the interval $[-k^*,k^*]$. The reason is that unlike the quench from the tilted N\'eel only pairs of quasiparticles traveling with 
opposite velocity are present. This means that at long times the asymmetry vanishes because the number of pairs in $A$ vanishes. 

We now discuss the quench from the tilted N\'eel state. Now, neither restoration of symmetry nor Mpemba effect occur in the 
absence of dissipation, although curves for different values of the tilting angle can cross at intermediate time. 
To write the crossing condition, let us 
notice that the leading order of the R\'enyi entanglement asymmetry at long time reads~\cite{ares2023lack} 
\begin{equation}
    \label{eq:asym_neel_tinf}
    \Delta S^{(n)}(\theta,t\to \infty)=\frac{1}{2} \ln \ell + \frac{1}{2}\ln\left(\frac{\pi n^{1/(n-1)}}{4} \int_0^{2\pi} \frac{dk}{2\pi} \frac{f_{11}(k,\theta)^2}{p_n(n_+(k,\theta),n_-(k,\theta))}\right)+ O(\ell^{1-n}),
\end{equation}
where $p_n$ is, in general, a cumbersome function of $n_\pm(k,\theta)$ (cf.~\eqref{eq:npm}), that also depends on the R\'enyi index $n$. For example, $p_2(n_+,n_-)=h_2(n_+)h_2(n_-)$ with $h_n$ defined in~\eqref{eq:hn}. For the following we do not need the explicit expression for $p_n$ (see  Ref.~\cite{ares2023lack}). By using~\eqref{eq:asym_neel_tinf} and~\eqref{eq:asym_leading_t0}, the crossing condition in Eq.~\eqref{eq:mpemba} becomes 
\begin{equation}
    \label{eq:mpemba_neel}
    \left\{
    \begin{gathered}
        \int_0^{2\pi}\frac{dk}{2\pi}\left(f_{11}(k,\theta_1)^2+f_{12}(k,\theta_1)^2\right)>\int_0^{2\pi}\frac{dk}{2\pi}\left(f_{11}(k,\theta_2)^2+f_{12}(k,\theta_2)^2\right),\\
        \int_0^{2\pi} \frac{dk}{2\pi} \frac{f_{11}(k,\theta_1)^2}{p_n(n_+(k,\theta_1),n_-(k,\theta_1))}<\int_0^{2\pi} \frac{dk}{2\pi} \frac{f_{11}(k,\theta_2)^2}{p_n(n_+(k,\theta_2),n_-(k,\theta_2))}.
    \end{gathered}
    \right.
\end{equation}
The conditions in~\eqref{eq:mpemba_neel} are quite similar to the ones in the presence of dissipation (cf.~\eqref{eq:mpemba_diss}), in contrast with the standard Mpemba conditions~\eqref{eq:mpemba_XY}. The first condition is the same, since at $t=0$ the dissipation has no effect, whereas the second one differs for the denominator $p_n$ in the integrand, which is absent in the 
dissipative case. 
The generalized quasiparticle picture discussed in Section~\ref{sec:chmom_quasipart} allows to interpret the crossing condition in Eq.~\eqref{eq:mpemba_neel}. To establish the condition~\eqref{eq:mpemba_neel} 
one has to compare the asymmetry in the initial state with that at long times. In the initial state, the 
asymmetry is proportional to the correlators $\langle c_k c_{2\pi-k}\rangle$ and $\langle c_k c_{\pi-k} \rangle$. Again, the former correlator is 
due to the members of the quadruplet traveling with opposite velocity. The latter one corresponds to the quasiparticles 
traveling at the same velocity, which are responsible for the lack of symmetry restoration at long times. 
The integral in the first inequality in Eq.~\eqref{eq:mpemba_neel} is the sum of the contributions coming from $\langle c_k c_{\pi-k}\rangle$ and $\langle c_k c_{2\pi-k}\rangle$ as it is clear from Eq.~\eqref{eq:cc_corr}. At the initial time all the quasiparticles in 
the quadruplet contribute to the asymmetry. 
On the other hand, the integral in the second inequality of Eq.~\eqref{eq:mpemba_neel} involves only 
$\langle c_k c_{\pi-k}\rangle$ because at long times, only the correlator $\langle c_k c_{\pi-k}\rangle$  appears. The correlator 
$\langle c_k c_{2\pi-k}\rangle$ vanishes at long times because the associated quasiparticles  have opposite 
velocities, and cannot be both in $A$ at the same time. The integrand 
in the second condition in~\eqref{eq:mpemba_neel} can be somehow identified as the ``asymmetry content'' of  
the pair $(k,\pi-k)$. Notice that  the normalization $p_n(n_+,n_-)$  depends on both the  correlator $\langle c_k c_{2\pi-k}\rangle$ and the correlator $\langle c_k^\dagger c_k \rangle$, 
via the functions $f_{11}(2k,\theta)$ and $g_{12}(2k,\theta)$ in Eq.~\eqref{eq:quasipart_neel_z2_2}. 
Summarizing, the crossing in the entanglement asymmetry occurs if initial states with larger asymmetry  
lead to  pairs $c_k c_{\pi-k}$ with smaller asymmetry.

The interpretation of condition~\eqref{eq:mpemba_diss} for the quench in the presence of  dissipation remains qualitatively the same. First, dissipation suppresses the correlations carried by the quadruplets at a rate independent of $k$. This means that 
the correlator $\braket{c_k c_{\pi-k}}$ vanishes at long times, implying symmetry restoration and the possibility of having 
Mpemba effect. 
Specifically, the condition  to observe the Mpemba effect is, again, that the initial state with larger total 
asymmetry leads to pairs $(k,\pi-k)$ with smaller asymmetry content, that at long times are destroyed by the dissipation at the same rate. Indeed, 
Eq.~\eqref{eq:mpemba_neel} and Eq.~\eqref{eq:mpemba_diss} are the same, except for the  factor, $p_n(n_+(k,\theta),n_-(k,\theta))$, which is dropped in Eq.~\eqref{eq:mpemba_diss}. The reason is that  it is subleading in $\lambda(t)$.
Notice that in the presence of dissipation, the entanglement asymmetry $\Delta S_A^{(n)}(\theta,t)$ decays exponentially with  time with a rate $2(\gamma_++\gamma_-)$, as can be seen from Eq.~\eqref{eq:asym_gpgm_S}. This is different in the unitary integrable cases studied so far, where the asymmetry  displays an algebraic decay in time (see Ref. \cite{rylands2023microscopic}). On the other hand, the classical Mpemba effect, which occurs, for instance, in colloidal systems~\cite{scharzendahl2022anomalous}, exhibits a similar exponential behaviour.
Finally, another similarity with the classical Mpemba effect is that in  the presence of dissipation, the steady state arising at long times does not depend on the initial state, in contrast with the quantum Mpemba effect.

\section{Numerical results}
\label{sec:numerics}

In this section we provide numerical evidence supporting our results. 
We focus on the quench from the tilted N\'eel state in the $XX$ chain because, as it was discussed, 
in the absence of dissipation 
the $U(1)$ symmetry of the $XX$ chain, which is broken by the initial state, is not restored at 
long times. Our results confirm that in the presence of dissipation the symmetry can be restored, and 
Mpemba effect can occur. The section is structured as follows. In subsection~\ref{sec:charged-num} we 
benchmark our results for the charged moments in the presence of gain and loss dissipation. In section~\ref{sec:asy-num} 
we focus on the Mpemba effect discussing the R\'enyi entanglement asymmetry.

\subsection{Charged moments}
\label{sec:charged-num}

\begin{figure}[t!]
\centering
    {\includegraphics[width=0.49\textwidth]{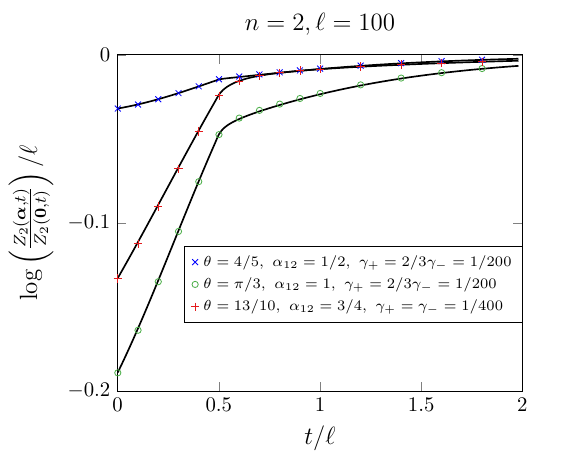}}
     {\includegraphics[width=0.49\textwidth]{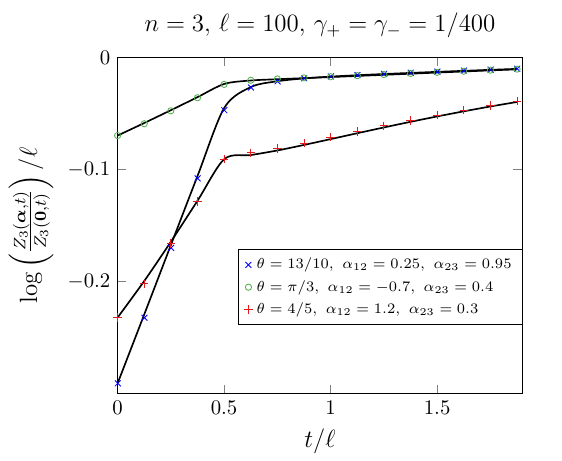}}
     \caption{Charged moments after the quench from the tilted Néel state as a function of time $t$ in a subsystem of size $\ell=100$ and different initial tilting angle $\theta$, gain/loss rates $\gamma_{\pm}$, and $\boldsymbol{\alpha}$ (cf.~\eqref{eq:chmom_general}). 
     , $n=2$ (left) or $n=3$ (right). The curves are Eqs.~\eqref{eq:chmom_general_diss}, while the symbols are the exact numerical results.}
     \label{fig:chmom}
\end{figure}

In Fig.~\ref{fig:chmom} we show numerical results for the charged moments 
$\ln(Z_n(\boldsymbol{\alpha},t)/Z_n(\boldsymbol{0},t))/\ell$ for 
$n=2$ and $n=3$ (left and right panel, respectively).
The charged moments 
are plotted against the rescaled time $t/\ell$, with $\ell$ the size of subsystem $A$. We show results in the weakly-dissipative 
hydrodynamic limit $t,\ell\to\infty$ with the ratio $t/\ell$ fixed and $\gamma^\pm\to0$ 
with fixed $\gamma^+\ell,\gamma^-\ell$. At fixed $\gamma^\pm$, in the limit $t\to\infty$ 
the charged moments exhibit an exponential relaxation to the steady state. In Fig.~\ref{fig:chmom} 
we consider several tilting  angles $\theta$, and values of $\alpha_{ij}:=\alpha_i-\alpha_j$. 
The symbols in Fig.~\ref{fig:chmom} are exact lattice results. 
The data are obtained by applying directly Eq.~\eqref{eq:chmom_log} to the real-space correlator $\Gamma(t)$ computed using the symbol in Eq.~\eqref{eq:G_diss}. 
The charged moments are always negative. 
At $\theta=0$
they vanish, whereas upon increasing $\theta$ they become larger in absolute value. 
The continuous lines in the Figure 
are obtained by using the quasiparticle picture (cf.~\eqref{eq:chmom_general_diss}). The agreement 
with the numerics is perfect.

\subsection{Entanglement asymmetry}
\label{sec:asy-num}

\begin{figure}[t!]
\centering
    {\includegraphics[width=0.49\textwidth]{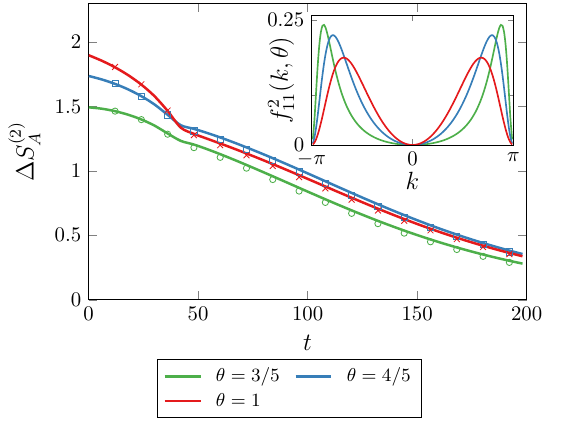}}
     {\includegraphics[width=0.49\textwidth]{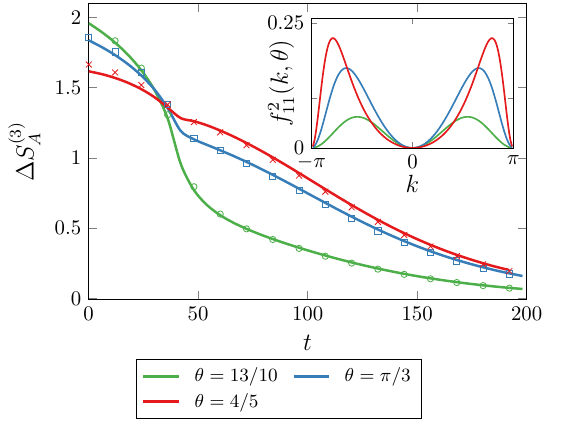}}
     \caption{R\'enyi entanglement asymmetry after the quench from the tilted Néel state in the $XX$ chain as a function of time $t$ for a subsystem size $\ell=80$, dissipation $\gamma_+=4\gamma_-=1/200$ (left panel) $ \gamma_+=\gamma_-=1/200$ (right panel) and for different initial tilting angle $\theta$. The curves have been obtained by performing the numerical Fourier transform of Eq.~\eqref{eq:chmom_general_diss}, while the symbols are the exact numerical values. The inset shows $f_{11}^2(k,\theta)$ as a 
     function of quasimomentum $k$. The vanishing behaviour at long times and the crossing at short one signal the quantum Mpemba effect. 
     } 
     \label{fig:ent_asymmn3}
\end{figure}

We now discuss the R\'enyi entanglement asymmetry. The entanglement asymmetry is obtained by 
plugging Eq.~\eqref{eq:chmom_general_diss} in~\eqref{eq:FT} and performing the Fourier transform numerically.  
The analytic prediction obtained by using the quasiparticle picture is reported with  the solid lines in Fig.~\ref{fig:ent_asymmn3}. We plot it for different tilting angles $\theta$, dissipation parameters $\gamma_+,\gamma_-$ and for $n=2,3$. Specifically, the left panel shows results for $n=2$ and $\gamma^+=4\gamma^-$. Fig.~\ref{fig:ent_asymmn3}  (right panel) plots results for $n=3$ and balanced gain and losses, i.e., for $\gamma^+=\gamma^-$. For $n=2$ we can also take advantage of the general analytical expression for the charged moments given in Eqs.~\eqref{eq:z2_generic_t0} and~\eqref{eq:z2_generic_tinf}. 
The symbols in the plots are exact numerical data. The agreement with the quasi-particle prediction is remarkable.

Fig.~\ref{fig:ent_asymmn3} shows that  there exist states that have larger asymmetry at time $t=0$ but faster restoration during the time evolution (for instance, the blue and red lines in the left panel). This is a clear signature of the quantum Mpemba effect. However, this behaviour is not generic,  meaning that having a larger tilting angle $\theta \in [0,\pi/2]$ does not automatically imply Mpemba. This is in contrast with  what happens for the quench from the tilted ferromagnetic state~\cite{ares2023entanglement}. The scenario 
depends dramatically on the dissipation. For instance, for balanced gain and losses (right panel in Fig.~\ref{fig:ent_asymmn3}) 
the Mpemba effect seems stronger.  In  the insets in Fig.~\ref{fig:ent_asymmn3} we show the condition ~\eqref{eq:mpemba_neel} for Mpemba effect. We plot $f_{11}^2(k,\theta)$ as a function of $k$. Clearly, when Mpemba occurs we observe that larger values of $\theta$, i.e., stronger  symmetry breaking 
in the initial state give rise to multiplets with smaller  asymmetry content, and smaller integral $\int dk f_{11}^2(k,\theta)$. This confirms the validity of~\eqref{eq:mpemba_neel}.  

\begin{figure}[t!]
\centering
    {\includegraphics[width=0.49\textwidth]{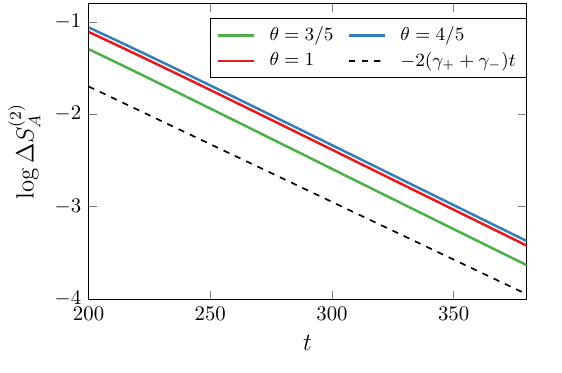}}
    {\includegraphics[width=0.49\textwidth]{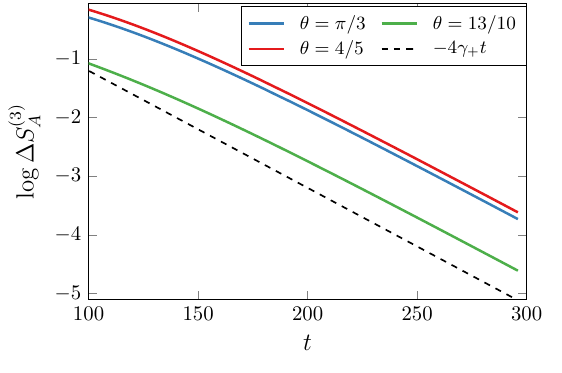}}
     \caption{Exponential decay in time of the entanglement asymmetry for different values of $\theta$, $n$, dissipation $\gamma_+=4\gamma_-=1/200$ (left panel) $ \gamma_+=\gamma_-=1/200$ (right panel)  and subsystem size $\ell=80$. The solid line correspond to the Fourier transform of Eq.~\eqref{eq:chmom_general_diss} plugged in Eqs.~\eqref{eq:FT} and~\eqref{eq:def_renyi}. The black dashed line is an eye-guide confirming the decay $\Delta S_A^{(n)}(\theta,t\to\infty)\sim e^{-2t(\gamma_++\gamma_-)}$ found in Eq.~\eqref{eq:asym_gpgm_S}.}
     \label{fig:ent_asymmn_exp}
\end{figure}

We finally check the asymptotic expansion of the entanglement asymmetry in Eq. \eqref{eq:asym_gpgm_S}. In Fig. \ref{fig:ent_asymmn_exp} we show the exact quasiparticle picture obtained by performing the Fourier transform of Eq.~\eqref{eq:chmom_general_diss} plugged in Eqs.~\eqref{eq:FT} and~\eqref{eq:def_renyi} (solid lines), and we prove that they exponentially decay as $\sim e^{-2t(\gamma_++\gamma_-)}$ for $n=2,3$ (dashed black lines).

\section{Conclusions}\label{sec:concl}

We investigated the quantum Mpemba effect in one-dimensional fermionic chains subject to dissipation. We employed the framework of 
Lindblad master equation~\cite{petruccione2002the}. First, by using the multidimensional stationary phase approximation we 
provided a fully rigorous derivation of quantum Mpemba effect for several quenches in free-fermion models in the absence 
of dissipation (see Ref.~\cite{ares2023entanglement} and~\cite{ares2023lack}). Based on these results we conjectured a formula 
for  the charged moments  in the hydrodynamic limit of long times and large intervals with their ratio fixed. Our formula exploits the fact that the generic quench produces entangled multiplets of excitations, rather than entangled pairs. Our results suggest that the framework of entangled multiplets is the natural one to formulate the quasiparticle picture interpretation for the dynamics of the entanglement asymmetry, and provides valuable insights also for quenches producing entangled pairs.  Then, we focused on the effects of gain and loss dissipation 
on the dynamics of the entanglement asymmetry. Specifically, we considered the quench from the tilted N\'eel state in the $XX$ chain. In the absence of dissipation the $U(1)$ symmetry of the model is not restored by the dynamics, and the quantum Mpemba effect does not occur. This is due to the presence of 
peculiar non-Abelian conservation laws of the $XX$ chain~\cite{ares2023lack,fagotti2014on}.  
Here we show that gain/loss dissipation can restore the symmetry and the quantum Mpemba effect can occur. Interestingly,  the condition that explains the effect is different with respect to the one studied in~\cite{rylands2023microscopic,murciano2023entanglement}. The reason can be traced back to the presence of entangled quadruplets of quasiparticles and not just entangled pairs. The correlator $\braket{c_kc_{\pi-k}}$ vanishes in the presence of dissipation, which explains why the symmetry is now restored. However, at long times the correlator gets contributions only from  quasiparticles traveling at the same velocity, since they  can be both in
$A$ at the same time. This is in contrast with quenches producing entangled pairs~\cite{murciano2023entanglement}. As these paired quasiparticles with the same velocity exist for any $k$, all the modes contribute to the  asymmetry at large time (cf. the second line of~\eqref{eq:mpemba_diss}), not just the ``slow'' ones, in contrast with the standard Mpemba effect~\cite{murciano2023entanglement}.

Let us now discuss some possible future directions. First, the quasiparticle picture that 
we proposed in Sec.~\ref{sec:chmom_quasipart} works for the charged moments. However,  our results suggest that 
it may be possible to extend the  approach to obtain directly the entanglement asymmetry. To do that, one has to 
first determine  the relationship between the integrands in Eqs.~\eqref{eq:asym_leading_t0}-\eqref{eq:asym_gpgm_S} and the fermionic correlation matrix in momentum space 
  (cf.~\eqref{eq:chmom_quasipart}).  
It would be also interesting to generalize our results for the charged moments and for the entanglement asymmetry  to other  types of Lindblad master equations, by exploiting the results of 
Ref.~\cite{carollo2022dissipative}. This would allow to better clarify the conditions under which 
``dissipative'' quantum Mpemba effect occurs. Furthermore, our results showed that the Mpemba effect, or the absence thereof,  is intertwined with the 
production of nontrivial entangled multiplets after the quench. It would be interesting to explore the effect of the size of the 
multiplets and of dissipation on the asymmetry, and  on the quantum Mpemba effect. To this purpose, one could exploit the
results of Ref.~\cite{caceffo2023negative}. Clearly, it would be interesting to go beyond the case of quadratic Lindblad equations, 
to study the effect of interactions. While it is  challenging to determine the entanglement asymmetry, by exploiting integrability it should be possible to study the effect of interactions and dissipation on the quantum Mpemba effect  
at the level of correlation functions~\cite{alba2023free}. 

\section*{Acknowledgments} 
We thank Filiberto Ares, Pasquale Calabrese and Vittorio Vitale for useful discussions and collaboration on a related topic. 
VA acknowledges support by the project “Artificially devised many-body quantum dynamics in low dimensions - ManyQLowD” funded by the MIUR Progetti di Ricerca di Rilevante Interesse Nazionale (PRIN) Bando 2022 - grant 2022R35ZBF.  SM thanks support from Caltech Institute for Quantum Information and Matter and the Walter Burke Institute for Theoretical Physics at Caltech. 

%\printbibliography
\section*{References}
\bibliographystyle{iopart-num.bst}
\bibliography{bibliography}

\providecommand{\newblock}{}
\begin{thebibliography}{10}
\expandafter\ifx\csname url\endcsname\relax
  \def\url#1{{\tt #1}}\fi
\expandafter\ifx\csname urlprefix\endcsname\relax\def\urlprefix{URL }\fi
\providecommand{\eprint}[2][]{\url{#2}}
% Bibliography created with iopart-num v2.1
% /biblio/bibtex/contrib/iopart-num

\bibitem{ares2023entanglement}
Ares F, Murciano S and Calabrese P 2023 {\em Nature Communications\/} {\bf 14}
  2036 ISSN 2041-1723
  \urlprefix\url{https://doi.org/10.1038/s41467-023-37747-8}

\bibitem{ares2023lack}
Ares F, Murciano S, Vernier E and Calabrese P 2023 {\em SciPost Phys.\/} {\bf
  15} 089 \urlprefix\url{https://scipost.org/10.21468/SciPostPhys.15.3.089}

\bibitem{capizzi2023entanglement}
Capizzi L and Mazzoni M 2023 {Entanglement asymmetry in the ordered phase of
  many-body systems: the Ising Field Theory} (\textit{Preprint}
  \eprint{arXiv:2307.12127})

\bibitem{capizzi2023universal}
Capizzi L and Vitale V 2023 {A universal formula for the entanglement asymmetry
  of matrix product states} (\textit{Preprint} \eprint{arXiv:2310.01962})

\bibitem{rylands2023microscopic}
Rylands C, Klobas K, Ares F, Calabrese P, Murciano S and Bertini B 2023
  (\textit{Preprint} \eprint{arXiv:2310.04419})

\bibitem{ferro2023nonequilibrium}
Ferro F, Ares F and Calabrese P {Non-equilibrium entanglement asymmetry for
  discrete groups: the example of the XY spin chain} (\textit{Preprint}
  \eprint{arXiv:2307.06902})

\bibitem{chen2023entanglement}
Chen M and Chen H~H 2023 {Entanglement asymmetry in 1+1-dimensional Conformal
  Field Theories} (\textit{Preprint} \eprint{arXiv:2310.15480})

\bibitem{ares2023page}
Ares F, Murciano S, Piroli L and Calabrese P 2023 An entanglement asymmetry
  study of black hole radiation (\textit{Preprint} \eprint{arXiv:2311.12683})

\bibitem{murciano2023entanglement}
Murciano S, Ares F, Klich I and Calabrese P 2024 {\em J. Stat. Mech.\/} {\bf
  2024} 013103 \urlprefix\url{https://dx.doi.org/10.1088/1742-5468/ad17b4}

\bibitem{khor2023confinement}
Khor B~J~J, Kürkçüoglu D~M, Hobbs T~J, Perdue G~N and Klich I 2023
  {Confinement and Kink Entanglement Asymmetry on a Quantum Ising Chain}
  (\textit{Preprint} \eprint{arXiv:2312.08601})

\bibitem{joshi2024observing}
Joshi L~K, Franke J, Rath A, Ares F, Murciano S, Kranzl F, Blatt R, Zoller P,
  Vermersch B, Calabrese P, Roos C~F and Joshi M~K 2024  (\textit{Preprint}
  \eprint{arXiv:2401.04270})

\bibitem{calabrese2016introduction}
Calabrese P, Essler F~H~L and Mussardo G 2016 {\em J. Stat. Mech.\/} {\bf 2016}
  064001 \urlprefix\url{https://doi.org/10.1088/1742-5468/2016/06/064001}

\bibitem{fagotti2014on}
Fagotti M 2014 {\em J. Stat. Mech.\/} {\bf 2014} P03016
  \urlprefix\url{https://dx.doi.org/10.1088/1742-5468/2014/03/P03016}

\bibitem{fagotti2014relaxation}
Fagotti M, Collura M, Essler F~H~L and Calabrese P 2014 {\em Phys. Rev. B\/}
  {\bf 89}(12) 125101
  \urlprefix\url{https://link.aps.org/doi/10.1103/PhysRevB.89.125101}

\bibitem{mpemba1969cool}
Mpemba E~B and Osborne D~G 1969 {\em Physics Education\/} {\bf 4} 172
  \urlprefix\url{https://dx.doi.org/10.1088/0031-9120/4/3/312}

\bibitem{ahn2016experimental}
Ahn Y~H, Kang H, Koh D~Y and Lee H 2016 {\em Korean Journal of Chemical
  Engineering\/} {\bf 33} 1903--1907 ISSN 1975-7220
  \urlprefix\url{https://doi.org/10.1007/s11814-016-0029-2}

\bibitem{chaddah2010overtaking}
Chaddah P, Dash S, Kumar K and Banerjee A 2010 Overtaking while approaching
  equilibrium (\textit{Preprint} \eprint{arXiv:1011.3598})

\bibitem{greaney2011mpemba}
Greaney P~A, Lani G, Cicero G and Grossman J~C 2011 {\em Metallurgical and
  Materials Transactions A\/} {\bf 42} 3907--3912 ISSN 1543-1940
  \urlprefix\url{https://doi.org/10.1007/s11661-011-0843-4}

\bibitem{lasanta2017when}
Lasanta A, Vega~Reyes F, Prados A and Santos A 2017 {\em Phys. Rev. Lett.\/}
  {\bf 119}(14) 148001
  \urlprefix\url{https://link.aps.org/doi/10.1103/PhysRevLett.119.148001}

\bibitem{kumar2020exponentially}
Kumar A and Bechhoefer J 2020 {\em Nature\/} {\bf 584} 64--68 ISSN 1476-4687
  \urlprefix\url{https://doi.org/10.1038/s41586-020-2560-x}

\bibitem{keller2018quenches}
Keller T, Torggler V, Jäger S~B, Schütz S, Ritsch H and Morigi G 2018 {\em
  New J. Phys.\/} {\bf 20} 025004
  \urlprefix\url{https://dx.doi.org/10.1088/1367-2630/aaa161}

\bibitem{Chatterjee2023}
Chatterjee A~K, Takada S and Hayakawa H 2023 {\em Phys. Rev. Lett.\/} {\bf 131}
  080402 \urlprefix\url{http://arxiv.org/abs/2304.02411
  https://link.aps.org/doi/10.1103/PhysRevLett.131.080402}

\bibitem{shapira2024mpemba}
Shapira S~A, Shapira Y, Markov J, Teza G, Akerman N, Raz O and Ozeri R 2024
  (\textit{Preprint} \eprint{arXiv:2401.05830})

\bibitem{zhang2024observation}
Zhang J, Xia G, Wu C~W, Chen T, Zhang Q, Xie Y, Su W~B, Wu W, Qiu C~W, xing
  Chen P, Li W, Jing H and Zhou Y~L 2024 Observation of quantum strong mpemba
  effect (\textit{Preprint} \eprint{arXiv:2401.15951})

\bibitem{wang2024mpemba}
Wang X and Wang J 2024 Mpemba effects in nonequilibrium open quantum systems
  (\textit{Preprint} \eprint{arXiv:2401.14259})

\bibitem{calabrese2005evolution}
Calabrese P and Cardy J 2005 {\em J. Stat. Mech.\/} {\bf 2005} P04010
  \urlprefix\url{https://doi.org/10.1088{\%}2F1742-5468{\%}2F2005{\%}2F04{\%}2Fp04010}

\bibitem{fagotti2008evolution}
Fagotti M and Calabrese P 2008 {\em Phys. Rev. A\/} {\bf 78}(1) 010306
  \urlprefix\url{https://link.aps.org/doi/10.1103/PhysRevA.78.010306}

\bibitem{alba2017entanglement}
Alba V and Calabrese P 2017 {\em Proceedings of the National Academy of
  Sciences\/} {\bf 114} 7947--7951 ISSN 0027-8424
  \urlprefix\url{https://www.pnas.org/content/114/30/7947}

\bibitem{calabrese2018entanglement}
Calabrese P 2018 {\em Physica A\/} {\bf 504} 31--44 ISSN 0378-4371 lecture
  Notes of the 14th International Summer School on Fundamental Problems in
  Statistical Physics
  \urlprefix\url{https://www.sciencedirect.com/science/article/pii/S037843711731018X}

\bibitem{bertini2018entanglementevolution}
Bertini B, Fagotti M, Piroli L and Calabrese P 2018 {\em J. Phys. A\/} {\bf 51}
  39LT01 \urlprefix\url{https://dx.doi.org/10.1088/1751-8121/aad82e}

\bibitem{bastianello2018spreading}
Bastianello A and Calabrese P 2018 {\em SciPost Phys.\/} {\bf 5} 033
  \urlprefix\url{https://scipost.org/10.21468/SciPostPhys.5.4.033}

\bibitem{caceffo2023negative}
Caceffo F and Alba V 2023 {\em Phys. Rev. B\/} {\bf 108}(13) 134434
  \urlprefix\url{https://link.aps.org/doi/10.1103/PhysRevB.108.134434}

\bibitem{nava2019lindblad}
Nava A and Fabrizio M 2019 {\em Phys. Rev. B\/} {\bf 100}(12) 125102
  \urlprefix\url{https://link.aps.org/doi/10.1103/PhysRevB.100.125102}

\bibitem{carollo2021exponentially}
Carollo F, Lasanta A and Lesanovsky I 2021 {\em Phys. Rev. Lett.\/} {\bf
  127}(6) 060401
  \urlprefix\url{https://link.aps.org/doi/10.1103/PhysRevLett.127.060401}

\bibitem{petruccione2002the}
Breuer H~P and Petruccione F 2002 {\em The theory of open quantum systems\/}
  (Great Clarendon Street: Oxford University Press)

\bibitem{bertini2018entanglement}
Bertini B, Tartaglia E and Calabrese P 2018 {\em J. Stat. Mech.\/} {\bf 2018}
  063104 \urlprefix\url{https://dx.doi.org/10.1088/1742-5468/aac73f}

\bibitem{peschel2009reduced}
Peschel I and Eisler V 2009 {\em J. Phys. A\/} {\bf 42} 504003
  \urlprefix\url{https://dx.doi.org/10.1088/1751-8113/42/50/504003}

\bibitem{prosen2008third}
Prosen T 2008 {\em New J. Phys.\/} {\bf 10} 043026
  \urlprefix\url{https://dx.doi.org/10.1088/1367-2630/10/4/043026}

\bibitem{alba2018entanglement}
Alba V and Calabrese P 2018 {\em SciPost Phys.\/} {\bf 4}(3) 17
  \urlprefix\url{https://scipost.org/10.21468/SciPostPhys.4.3.017}

\bibitem{alba2019quantum}
Alba V and Calabrese P 2019 {\em {EPL} (Europhysics Letters)\/} {\bf 126} 60001
  \urlprefix\url{https://doi.org/10.1209/0295-5075/126/60001}

\bibitem{alba2018entanglementand}
Alba V 2018 {\em Phys. Rev. B\/} {\bf 97}(24) 245135
  \urlprefix\url{https://link.aps.org/doi/10.1103/PhysRevB.97.245135}

\bibitem{mestyan2020molecular}
Mestyán M and Alba V 2020 {\em SciPost Phys.\/} {\bf 8}(4) 55
  \urlprefix\url{https://scipost.org/10.21468/SciPostPhys.8.4.055}

\bibitem{alba2021generalized}
Alba V, Bertini B, Fagotti M, Piroli L and Ruggiero P 2021 {\em J. Stat.
  Mech.\/} {\bf 2021} 114004
  \urlprefix\url{https://doi.org/10.1088/1742-5468/ac257d}

\bibitem{alba2021spreading}
Alba V and Carollo F 2021 {\em Phys. Rev. B\/} {\bf 103}(2) L020302
  \urlprefix\url{https://link.aps.org/doi/10.1103/PhysRevB.103.L020302}

\bibitem{carollo2022dissipative}
Carollo F and Alba V 2022 {\em Phys. Rev. B\/} {\bf 105}(14) 144305
  \urlprefix\url{https://link.aps.org/doi/10.1103/PhysRevB.105.144305}

\bibitem{alba2022hydrodynamics}
Alba V and Carollo F 2022 {\em J. Phys. A\/} {\bf 55} 74002
  \urlprefix\url{https://doi.org/10.1088/1751-8121/ac48ec}

\bibitem{alba2022logarithmic}
Alba V and Carollo F 2022  \urlprefix\url{https://arxiv.org/abs/2205.02139}

\bibitem{klobas2021entanglement}
Klobas K and Bertini B 2021 {\em SciPost Phys.\/} {\bf 11}(6) 107
  \urlprefix\url{https://scipost.org/10.21468/SciPostPhys.11.6.107}

\bibitem{klobas2021exact}
Klobas K, Bertini B and Piroli L 2021 {\em Phys. Rev. Lett.\/} {\bf 126}(16)
  160602
  \urlprefix\url{https://link.aps.org/doi/10.1103/PhysRevLett.126.160602}

\bibitem{bertini2022growth}
Bertini B, Klobas K, Alba V, Lagnese G and Calabrese P 2022 {\em Phys. Rev.
  X\/} {\bf 12}(3) 031016
  \urlprefix\url{https://link.aps.org/doi/10.1103/PhysRevX.12.031016}

\bibitem{alba2017quench}
Alba V and Calabrese P 2017 {\em Phys. Rev. B\/} {\bf 96}(11) 115421
  \urlprefix\url{https://link.aps.org/doi/10.1103/PhysRevB.96.115421}

\bibitem{alba2017renyi}
Alba V and Calabrese P 2017 {\em J. Stat. Mech.\/} {\bf 2017} 113105
  \urlprefix\url{https://doi.org/10.1088/1742-5468/aa934c}

\bibitem{mestyan2018renyi}
Mesty{\'{a}}n M, Alba V and Calabrese P 2018 {\em J. Stat. Mech.\/} {\bf 2018}
  083104 \urlprefix\url{https://doi.org/10.1088/1742-5468/aad6b9}

\bibitem{yamashika2023time}
Yamashika S, Ares F and Calabrese P 2023  (\textit{Preprint}
  \eprint{arXiv:2310.18160})

\bibitem{gibbins2023quench}
Gibbins M, Jafarizadeh A, Smith A and Bertini B 2023  (\textit{Preprint}
  \eprint{arXiv:2310.18227})

\bibitem{scharzendahl2022anomalous}
Schwarzendahl F~J and L\"owen H 2022 {\em Phys. Rev. Lett.\/} {\bf 129}(13)
  138002
  \urlprefix\url{https://link.aps.org/doi/10.1103/PhysRevLett.129.138002}

\bibitem{alba2023free}
Alba V 2023  (\textit{Preprint} \eprint{arXiv:2309.12978})

\end{thebibliography}

\begin{appendices}

\section{Proof of equations~\eqref{eq:nodiss_t0} and~\eqref{eq:nodiss_tinf}}
\label{app:proof_nodiss}

Let us start from the small-time term~\eqref{eq:nodiss_t0}. We have to determine the behavior 
in the large $\ell$ limit of 
\begin{equation}
    \label{eq:dets}
    \det\left(\frac{\mathds{1}-\gcal_0(k)}{2}\right)^n\det\left(\mathds{1}+\mathcal{W}_0(k)\right),
\end{equation}
where $\mathcal{W}_0(k)=\prod_{j=1}^{n}(\mathds{1}+\gcal_0(k))(\mathds{1}-\gcal_0(k))^{-1}e^{i(\alpha_j-\alpha_{j+1})n_A}$, and 
$\gcal_0(k)$ is defined in~\eqref{eq:G_pm0}. First, we notice that $\gcal^2_0(k)=\mathds{1}$, and the eigenvalues of $\gcal_0(k)$ are $\pm1$, each one being doubly degenerate. Thus $\mathds{1}-\gcal_0(k)$ is not invertible. However, for each diverging term in  $(\mathds{1}-\gcal_0(k))^{-1}$, there is a correspondent vanishing factor in the first determinant in~\eqref{eq:dets}, which 
heals the singular behaviour. To proceed, we replace $\gcal_0$ with $c\gcal_0$, where $c$ is a number, and then take the limit $c\to1$. We easily find 
\begin{equation}
    \label{eq:det1}
    \det\left(\frac{\mathds{1}-c\gcal_0(k)}{2}\right)^n=\frac{(1-c^2)^{2n}}{16^n}.
\end{equation}
Now, a crucial fact is that since  $\mathcal{W}_0(k)$ a symplectic matrix, if $\lambda$ is an eigenvalue, then $1/\lambda$ is also 
an eigenvalue. This implies, by simple algebra 
\begin{equation}
    \label{eq:det_to_tr}
    \det\left(\mathds{1}+\mathcal{W}_0(k)\right)= 2+2 \mathrm{Tr}\left[\mathcal{W}_0(k)\right]+\frac{1}{2}\left(\mathrm{Tr}\left[\mathcal{W}_0(k)\right]^2-\mathrm{Tr}\left[\mathcal{W}^2_0(k)\right]\right).
\end{equation}
It is also straightforward to check that $(\mathds{1}+c\gcal_0(k))(\mathds{1}-c\gcal_0(k))^{-1}=((1+c^2)\mathds{1}+2c\gcal_0(k))/(1-c^2)$. Now,  only the last two terms in~\eqref{eq:det_to_tr} are singular enough to survive in the limit $c\to1$ when multiplying~\eqref{eq:det1} and~\eqref{eq:det_to_tr}. In summary, taking the limit $c\to1$ we obtain:
\begin{equation}
    \label{eq:cto1}
    \det\left(\frac{\mathds{1}-\gcal_0(k)}{2}\right)^n\det\left(\mathds{1}+\mathcal{W}_0(k)\right)=\frac{\mathrm{Tr}\left[\overline{\mathcal{W}_0(k)}\right]^2-\mathrm{Tr}\left[\overline{\mathcal{W}_0(k)}^2\right]}{2\cdot 4^n},
\end{equation}
where $\overline{\mathcal{W}_0(k)}=\prod_{j=1}^{n}(\mathds{1}+\gcal_0(k))e^{i(\alpha_j-\alpha_{j+1})n_A}$ is the 
same as $\mathcal{W}_0(k)$ except for the diverging denominator. We now have to evaluate the two remaining traces. Again, it is 
convenient to first evaluate the ``moments" $M_\mathbf{h}^{(0)}$ obtained by Taylor-expanding the complex exponentials 
in~\eqref{eq:cto1} as 
\begin{equation}
    \label{eq:n_moments}
    M_\mathbf{h}^{(0)}:=\mathrm{Tr}\left[(\mathds{1}+\gcal_0(k))n_A^{h_1}...(\mathds{1}+\gcal_0(k))n_A^{h_L}\right],
\end{equation}
and then exploit the analiticity in $n_A$ of $\overline{\mathcal{W}_0(k)}$. 
In~\eqref{eq:n_moments} it is meant that each term $n_A$ is multiplyied by a factor $i(\alpha_j-\alpha_{j+1})$, and there is 
a  factor $1/(h_j!)$. 
For now, we neglect this fact, and we will restore the dependence on $\alpha_j$ at the end. 
We notice that since $(\mathds{1}+\gcal_0(k))^2=2(\mathds{1}+\gcal_0(k))$,  we can  rewrite the moments~\eqref{eq:n_moments} as
\begin{equation}
    \label{eq:n_moments_2}
    M_\mathbf{h}^{(0)}=2^{\#\text{ even }h_j}\mathrm{Tr}\left[(n_A(\mathds{1}+\gcal_0(k)))^{\#\text{ odd }h_j}\right]. 
\end{equation}
The matrix $n_A(\mathds{1}+\gcal_0(k))$ has only two opposite nonzero eigenvalues, $\pm2\sqrt{1-m_N(k,\theta)^2}$, so we have 
\begin{equation}
    \label{eq:n_moments_3}
    M_\mathbf{h}^{(0)}=2^L\left(\sqrt{1-m_N(k,\theta)^2}^{\sum_{j=1}^L \frac{1-(-1)^{h_j}}{2}}+\left(-\sqrt{1-m_N(k,\theta)^2}\right)^{\sum_{j=1}^L \frac{1-(-1)^{h_j}}{2}}\right).
\end{equation}
If we now resum the series for the exponentials of $n_A$, after restoring the dependence on $\alpha_j$, the two terms in~\eqref{eq:n_moments_3}  yield $\cos\alpha_{j,j+1}\pm i \sin\alpha_{j,j+1}\sqrt{1-m_N(k,\theta)^2}$ where the plus and the minus sign correspond to  the first and the second term in~\eqref{eq:n_moments_3}, respectively. In conclusion, we obtain 
\begin{multline}
    \label{eq:tracesW0}
        \mathrm{Tr}\left[\overline{\mathcal{W}_0(k)}\right]=2^n\Big(\prod_{j=1}^n (\cos(\alpha_{j,j+1}) + i \sin(\alpha_{j,j+1})\sqrt{1-m_N(k,\theta)^2})\\
        +\prod_{j=1}^n \left(\cos(\alpha_{j,j+1}) - i \sin(\alpha_{j,j+1})\sqrt{1-m_N(k,\theta)^2}\right)\Big),
        \end{multline}
        and
        \begin{multline}
        \label{eq:tracesW0-1}
        \mathrm{Tr}\left[\overline{\mathcal{W}_0(k)}^2\right]=2^{2n}\left(\prod_{j=1}^n \left(\cos(\alpha_{j,j+1}) + i \sin(\alpha_{j,j+1})\sqrt{1-m_N(k,\theta)^2}\right)^2+\right.\\
        \left.+\prod_{j=1}^n \left(\cos(\alpha_{j,j+1}) - i \sin(\alpha_{j,j+1}) \sqrt{1-m_N(k,\theta)^2}\right)^2 \right).
\end{multline}
By substituting back~\eqref{eq:tracesW0} and~\eqref{eq:tracesW0-1} in~\eqref{eq:cto1}, we finally obtain 
\begin{multline}
    \label{eq:nodiss_t0_final}
        \det\left(\mathds{1}+\mathcal{W}_0(k)\right)\det\left(\mathds{1}+\mathcal{W}_0(k)\right)=\prod_{j=1}^n \left(\cos^2\alpha_{j,j+1} + \sin^2\alpha_{j,j+1}\left(1-m_N(k,\theta)^2\right)\right)=\\
        =\prod_{j=1}^n \left(1- m_N(k,\theta)^2\sin^2\alpha_{j,j+1}\right),
\end{multline}
which gives exactly the expected expression~\eqref{eq:nodiss_t0}.

Let us now focus on~\eqref{eq:nodiss_tinf}. We have to calculate compute 
\begin{equation}
    \label{eq:dets_tinf}
    \det\left(\frac{\mathds{1}-\gcal_\infty(k)}{2}\right)^n\det\left(\mathds{1}+\mathcal{W}_\infty(k)\right),
\end{equation}
where $\mathcal{W}_\infty(k)=\prod_{j=1}^{n}(\mathds{1}+\gcal_\infty(k))(\mathds{1}-\gcal_\infty(k))^{-1}e^{i(\alpha_j-\alpha_{j+1})n_A}$. The first determinant is readily obtained from the eigenvalues of $\gcal_\infty$, which are $\pm g_{12}(k,\theta) \pm f_{11}(k,\theta)$ (all 
the four possible sign choices are allowed), as 
\begin{equation}
    \label{eq:det1_tinf}
    \det\left(\frac{\mathds{1}-\gcal_\infty(k)}{2}\right)^n= \left(n_+(k,\theta)\left(1-n_+(k,\theta)\right)n_-(k,\theta)\left(1-n_-(k,\theta)\right)\right)^n.
\end{equation}
As for the second determinant in~\eqref{eq:dets_tinf}, we notice that if 
we rewrite it in the eigenbasis of $\gcal_\infty(k)$, we obtain 
\begin{multline}
    \label{eq:det2_tinf-1}
        \det\left(\mathds{1}+\mathcal{W}_\infty(k)\right)=\det\left[\mathds{1}+\begin{pmatrix}
        p&0&0&0\\0&q&0&0\\0&0&p^{-1}&0\\0&0&0&q^{-1}
    \end{pmatrix}\exp\left(i\alpha_{1,2}\begin{pmatrix}
        \sigma_x&0_{2\times2}\\0_{2\times2}&\sigma_x
    \end{pmatrix}\right)...\right.\\
    \left....\begin{pmatrix}
        p&0&0&0\\0&q&0&0\\0&0&p^{-1}&0\\0&0&0&q^{-1}
    \end{pmatrix}\exp\left(i\alpha_{n,1}\begin{pmatrix}
        \sigma_x&0_{2\times2}\\0_{2\times2}&\sigma_x
    \end{pmatrix}\right)\right]
\end{multline}
where  we defined $p:=n_+(k,\theta)/(1-n_+(k,\theta))$, $q:=n_-(k,\theta)/(1-n_-(k,\theta))$, and $\sigma_x$ is the Pauli matrix. The block-diagonal form of~\eqref{eq:det2_tinf-1} implies that the determinant can be split as the product of the determinants of the two $2\times2$ blocks. Each $2\times 2$ block $\mathcal{W}^{(\rho)}_\infty(k)$, with $\rho=\uparrow$ and $\rho=\downarrow$ denoting the 
upper-left and lower-right block in~\eqref{eq:det2_tinf-1}, respectively, can be written as 
\begin{equation}
    \label{eq:det_to_tr_tinf}
    \det\left(\mathds{1}+\mathcal{W}^{(\rho)}_\infty(k)\right)=1+\mathrm{Tr}\left[\mathcal{W}^{(\rho)}_\infty(k)\right]+\det \mathcal{W}^{(\rho)}_\infty(k),
\end{equation}
It is straightforward  to check that $\det(\mathcal{W}^{(\uparrow)}_\infty(k))=(pq)^n$, $\det(\mathcal{W}^{(\downarrow)}_\infty(k))=(pq)^{-n}$, and $\mathrm{Tr}[\mathcal{W}^{(\downarrow)}_\infty(k)]=(pq)^{-n}\;\mathrm{Tr}[\mathcal{W}^{(\uparrow)}_\infty(k)]$. Thus, the only term to evaluate in~\eqref{eq:det_to_tr_tinf} is $\mathrm{Tr}[\mathcal{W}^{(\uparrow)}_\infty(k)]$. Again, it is useful to first calculate the moments
\begin{equation}
    \label{eq:n_moments_tinf}
    M_\mathbf{h}^{(\infty)}:=\mathrm{Tr}\left[\begin{pmatrix}
        p&0\\0&q
    \end{pmatrix}\sigma_x^{h_1}...\begin{pmatrix}
        p&0\\0&q
    \end{pmatrix}\sigma_x^{h_L}\right].
\end{equation}
We now split the diagonal matrix as $(p+q) \mathds{1}/2+(p-q) \sigma_z/2$ and exploit the 
anticommutation relation $\{\sigma_x,\sigma_z\}=0$ to write:
\begin{equation}
    \label{eq:n_moments_tinf_intermediate}
        M_\mathbf{h}^{(\infty)}=\mathrm{Tr}\left\{\prod_{j=0}^{L-1}\left[\left(\frac{p+q}{2} \mathds{1}+(-1)^{\sum_{k=1}^j h_k}\frac{p-q}{2}\sigma_z\right)\right]\sigma_x^{\sum_{j=1}^{L}h_j}\right\}.
\end{equation}
After performing the trace, one obtains 
\begin{multline}
\label{eq:n_moments_tinf_intermediate-2}
        M_\mathbf{h}^{(\infty)}=2^{-L}\prod_{j=0}^{L-1} \left((p+q)+(-1)^{\sum_{k=1}^j h_k}(p-q)\right)\\+2^{-L}\prod_{j=0}^{L-1} \left((p+q)-(-1)^{\sum_{k=1}^j h_k}(p-q)\right)\frac{1+(-1)^{\sum_{j=1}^L h_j}}{2}.
\end{multline}
%\begin{equation}
%    \label{eq:n_moments_tinf_intermediate}
%    \begin{gathered}
%        M_\mathbf{h}^{(\infty)}=\mathrm{Tr}\left[\left(\frac{p+q}{2} \mathds{1}+\frac{p-q}{2}\sigma_z\right)\left(\frac{p+q}{2} \mathds{1}+(-1)^{h_1}\frac{p-q}{2}\sigma_z\right)\times\right.\\
%        \left. \times \left(\frac{p+q}{2} \mathds{1}+(-1)^{h_1+h_2}\frac{p-q}{2}\sigma_z\right)...\left(\frac{p+q}{2} \mathds{1}+(-1)^{\sum_{j=1}^{L-1}h_j}\frac{p-q}{2}\sigma_z\right)\sigma_x^{\sum_{j=1}^{L}h_j}\right]=\\
%        =2^{-L}\left( \left((p+q)+(p-q)\right)\left((p+q)+(-1)^{h_1}(p-q)\right)...\left((p+q)+(-1)^{\sum_{j=1}^{L-1}h_j}(p-q)\right)+\right.\\
%        \left.+\left((p+q)-(p-q)\right)\left((p+q)-(-1)^{h_1}(p-q)\right)...\left((p+q)-(-1)^{\sum_{j=1}^{L-1}h_j}(p-q)\right)\right)\frac{1+(-1)^{\sum_{j=1}^L h_j}}{2}.
%    \end{gathered}
%\end{equation}
%
At this point, we notice that if we expand the two products in~\eqref{eq:n_moments_tinf_intermediate_2}, they 
give opposite terms every time we choose an odd number of factors $p-q$, and equal ones when we choose an even number. 
Thus we have
\begin{multline}
    \label{eq:n_moments_tinf_intermediate_2}
        M_\mathbf{h}^{(\infty)}=2^{1-L}\frac{1+(-1)^{\sum_{i=1}^L h_i}}{2} \sum_{j=0}^{\lfloor L/2\rfloor}{\Bigg(} (p+q)^{L-2j} (p-q)^{2j}\\
        \left.\sum_{0\leq x_1<x_2<...<x_{2j}\leq L-1} (-1)^{\sum_{i=1}^{x_1}h_i}\cdot...\cdot(-1)^{\sum_{i=1}^{x_{2j}}h_i}\right)=\\
        =2^{1-L}\frac{1+(-1)^{\sum_{i=1}^L h_i}}{2}\sum_{j=0}^{\lfloor L/2\rfloor}{\Bigg(} (p+q)^{L-2j} (p-q)^{2j}\\
        \left.\sum_{0\leq x_1<x_2<...<x_{2j}\leq L-1} (-1)^{\sum_{i=x_1+1}^{x_2}h_i}(-1)^{\sum_{i=x_2+1}^{x_3}h_i}\cdot...\cdot(-1)^{\sum_{i=x_{2j}+1}^{L}h_i}\right). 
\end{multline}
By grouping the $h_j$ in such a way that in each product none of them appears more than once, 
Eq.~\eqref{eq:n_moments_tinf_intermediate_2} 
becomes 
\begin{multline}
    \label{eq:n_moments_tinf_final}
        M_\mathbf{h}^{(\infty)}=2^{-L}\sum_{j=0}^{\lfloor L/2\rfloor}{\Bigg(} (p+q)^{L-2j} (p-q)^{2j}\cdot\\
        \cdot\left(\sum_{0\leq x_1<x_2<...<x_{2j}\leq L-1} (-1)^{\sum_{i=x_1+1}^{x_2}h_i}\cdot...\cdot(-1)^{\sum_{i=x_{2j-1}+1}^{x_{2j}}h_i}+\right.\\
        \left.\left.+\sum_{0\leq x_1<x_2<...<x_{2j}\leq L-1} (-1)^{\sum_{i=1}^{x_1}h_i}\cdot...\cdot(-1)^{\sum_{i=x_{2j-1}+1}^{x_{2j}}h_i}\right)\right).
\end{multline}
We can now sum over $h_j$, restoring the dependence on the $\alpha_j$.  We obtain
\begin{multline}
    \label{eq:trace_tinf_intermediate}
        \mathrm{Tr}\left[\mathcal{W}_\infty(k)\right]=2^{-n}\sum_{j=0}^{\lfloor n/2\rfloor}{\Bigg(} (p+q)^{n-2j} (p-q)^{2j}\\
        \times\left(\sum_{0\leq x_1<x_2<...<x_{2j}\leq n-1} \exp\left(2i\left(\alpha_{x_1+1}-\alpha_{x_2+1}+...+\alpha_{x_{2j-1}+1}-\alpha_{x_{2j}+1}\right)\right)+\right.\\
        \left.\left.+\sum_{0\leq x_1<x_2<...<x_{2j}\leq n-1} \exp\left(-2i\left(\alpha_{x_1+1}-\alpha_{x_2+1}+...+\alpha_{x_{2j-1}+1}-\alpha_{x_{2j}+1}\right)\right)\right)\right). 
\end{multline}
Eq.~\eqref{eq:trace_tinf_intermediate} becomes
\begin{multline}
    \label{eq:trace_tinf_intermediate_2}
        \mathrm{Tr}\left[\mathcal{W}_\infty(k)\right]=2^{1-n}\sum_{j=0}^{\lfloor n/2\rfloor}{\Bigg(} (p+q)^{n-2j} (p-q)^{2j}\\
        \left.\times \sum_{1\leq x_1<x_2<...<x_{2j}\leq n}\left(1-2\sin^2\left(\alpha_{x_1}-\alpha_{x_2}+...+\alpha_{x_{2j-1}}-\alpha_{x_{2j}}\right)\right)\right). 
\end{multline}
Eq.~\eqref{eq:trace_tinf_intermediate_2} can be rewritten as 
\begin{equation}
    \label{eq:trace_tinf_final}
        \mathrm{Tr}\left[\mathcal{W}_\infty(k)\right]=p^n+q^n-\frac{f_{11}(k,\theta)^2}{(1-n_+(k,\theta))^n(1-n_-(k,\theta))^n}\tilde{h}_n(\boldsymbol{\alpha},k,\theta), 
\end{equation}
where $\tilde h_n(\boldsymbol{\alpha},k,\theta)$ and $n_\pm(k,\theta)$ are the same as in~\eqref{eq:nodiss_tinf}. 
Eq.~\eqref{eq:trace_tinf_final} implies 
\begin{equation}
    \label{eq:det2_tinf_final}
    \begin{gathered}
        \det\left(\mathds{1}+\mathcal{W}_\infty(k)\right)=\left(1+ p^n+q^n-\frac{f_{11}(k,\theta)^2}{(1-n_+(k,\theta))^n(1-n_-(k,\theta))^n}\tilde{h}_n(\boldsymbol{\alpha},k,\theta) + (pq)^n\right)\times\\
        \times \left(1+p^{-n}+q^{-n}-\frac{f_{11}(k,\theta)^2}{n_+(k,\theta)^n n_-(k,\theta)^n}\tilde{h}_n(\boldsymbol{\alpha},k,\theta) + (pq)^{-n}\right),
    \end{gathered}
\end{equation}
and substituting everything back into~\eqref{eq:dets_tinf}, we finally obtain:
\begin{equation}
    \label{eq:dets_tinf_subs}
    \det\left(\frac{\mathds{1}-\gcal_\infty(k)}{2}\right)^n\det\left(\mathds{1}+\mathcal{W}_\infty(k)\right)=\left(h_n(n_+(k,\theta)h_n(n_-(k,\theta))-f_{11}(k,\theta)^2\tilde{h}_n(\boldsymbol{\alpha},k,\theta)\right)^2,
\end{equation}
which gives exactly the expected expression~\eqref{eq:nodiss_tinf}.

\subsection{Considerations on the dissipative case}\label{app:diss}

In the presence of gain and loss dissipation, it is challenging to evaluate  the two determinants in~\eqref{eq:chmom_general_diss}, in contrast 
with the non dissipative case (see Appendix~\ref{app:proof_nodiss}). The reason is that the Fourier transform $\gcal'(k)$ of the fermionic correlator, defined in~\eqref{eq:G_diss}, loses some crucial properties that we used in the non-dissipative case. Indeed, it is no longer true that $(\gcal'_0(k,\theta,t))^2=\mathds{1}$ (cf.~\eqref{eq:G'_pm0}), which  we heavily exploited to calculate the short-time term $\ln(\det\mathcal{Z}_{n,0}^{(\boldsymbol{\alpha})}(k,t))$. Moreover,  the generalization of~\eqref{eq:det1_tinf} exhibits a block-diagonal  structure  in the eigenbasis of $\gcal'_\infty$, although it is more complicated (cf.~\eqref{eq:det2_tinf}). 
In the following we show that for generic dissipation rates $\gamma^\pm$ we are able to determine analytically $\ln(\det \mathcal{Z'}_{n,0}^{(\boldsymbol{\alpha})})$ and $\ln(\det \mathcal{Z'}_{n,\infty}^{(\boldsymbol{\alpha})})$ only for 
small values of $n$. For balanced gain and losses, i.e., $\gamma^+=\gamma^-$ we provide analytic results for $\ln(\det \mathcal{Z'}_{n,\infty}^{(\boldsymbol{\alpha})})$ for arbitrary $n$, whereas we provide a closed-form expression for $\ln(\det \mathcal{Z'}_{n,0}^{(\boldsymbol{\alpha})})$ only for $n=2$.  

We start discuss the case of balanced gain and losses. For  $\gamma_+=\gamma_-$ the second term in~\eqref{eq:G_diss} disappears, since we have $n_\infty=1/2$. Thus, we have  $\gcal'_t=\lambda(t)\gcal_t$, where we define $\lambda(t):=e^{-(\gamma_++\gamma_-)t}$. The derivation of $\ln(\det\mathcal{Z}_{n,\infty}^{\boldsymbol{(\alpha)}})$ discussed in  the previous section can be adapted to the dissipative case. The result is  the same result as in  the non-dissipative case, but with the substitutions $g_{12}(k,\theta)\to\lambda(t)g_{12}(k,\theta)$, $f_{11}(k,\theta)\to\lambda(t)f_{11}(k,\theta)$. We obtain 
\begin{equation}
    \label{eq:zn_gpgm_tinf_app}
    \ln( \det \mathcal{Z'}_{n,\infty}^{(\boldsymbol{\alpha})} )= 2 \ln \left( h_n(n'_+(k,\theta,t))h_n(n'_-(k,\theta,t))-\lambda(t)^2 f_{11}(k,\theta)^2 \tilde{h}'_n(\boldsymbol{\alpha},k,\theta,t) \right),
\end{equation}
where $n'_\pm (k,\theta,t):=(1+\lambda(t)g_{12}(k,\theta)\pm\lambda(t)f_{11}(k,\theta))/2$ and $\tilde{h}'_n(\boldsymbol{\alpha},k,\theta,t)$ is obtained by replacing $f_{11}(k,\theta)\rightarrow \lambda(t)f_{11}(k,\theta)$ and $n_\pm(k,\theta) \rightarrow n'_\pm (k,\theta,t)$ in Eq.~\eqref{eq:angles_tinf}. 

To determine $\ln(\det \mathcal{Z'}_{n,0}^{(\boldsymbol{\alpha})})$, since now $\gcal'(k,\theta,t)^2=\lambda(t)^2\mathds{1}$, one would need 
to evaluate the moments ${M'}_{\mathbf{h}}^{(0)}$ (see~\eqref{eq:dets} for the non-dissipative case ) as 
\begin{multline}
    \label{eq:n_moments'_t0}
    {M'}_\mathbf{h}^{(0)}:=\mathrm{Tr}\left[\prod_{j=1}^L\frac{\mathds{1}+\gcal'_0(k,t)}{\mathds{1}-\gcal'_0(k,t)}n_A^{h_j}\right]=\\
    =2^{-L}\;\mathrm{Tr}\left[\prod_{j=1}^L\left(\frac{1+\lambda(t)}{1-\lambda(t)}\left(\mathds{1}+\gcal_0(k)\right)+\frac{1-\lambda(t)}{1+\lambda(t)}\left(\mathds{1}-\gcal_0(k)\right)\right)n_A^{h_j}\right]. 
\end{multline}
The evaluation of the trace in~\eqref{eq:n_moments'_t0} is not straightforward. Indeed, notice that for 
even $h_j$,  Eq.~\eqref{eq:n_moments'_t0} is more complicated than in the non-dissipative case (cf.~\eqref{eq:n_moments}). 
 Anyway, we can evaluate the moments~\eqref{eq:n_moments'_t0} for small values of $L$, obtaining  
 a closed formula for $\ln( \det \mathcal{Z'}_{n,\infty}^{(\boldsymbol{\alpha})})$ for small values of $n$, 
 via Eq.~\eqref{eq:det_to_tr}. For example, for $n=2$ we obtain
\begin{equation}
    \label{eq:z2_gpgm_t0_app}
    \ln\left(\det \mathcal{Z'}_{2,0}^{(\boldsymbol{\alpha})}\right)=2 \ln \left(\frac{(1+\lambda^2)^2}{4} -\lambda^2 m_N(k,\theta)^2\sin^2(\alpha_{1,2})\right). 
\end{equation}

 We then know the leading order in the hydrodynamic limit of the charged moments with small $n$ in a completely analytic form, by substituting~\eqref{eq:z2_gpgm_t0_app}  and~\eqref{eq:zn_gpgm_tinf_app} into~\eqref{eq:chmom_general_diss}.

 %In Figure~\ref{fig:chmom}, we compare our formulas to numerical data, obtained by applying directly~\eqref{eq:chmom_log} to the real-space correlator $\Gamma(t)$. The agreement is striking, already for the relatively small subsystem size $\ell=100$.\\
For generic $\gamma_+,\gamma_-$, the short-time term $\ln(\det \mathcal{Z'}_{n,0}^{(\boldsymbol{\alpha})})$ is challenging to 
compute.  The result for $n=2$, which is obtained by employing~\eqref{eq:det_to_tr}, is already very cumbersome, and it reads 
\begin{multline}
    \label{eq:z2_generic_t0_app}
        \ln\left(\det \mathcal{Z'}_{2,0}^{(\boldsymbol{\alpha})}\right)=\ln{\Bigg[}\lambda^2(m_N(k,\theta)^2-1)s^2((\lambda -1)^3 s^2+(\lambda-1)(\lambda^2+1))^2 + \\
        +\left(\frac{(1+\lambda^2)^2+2(\lambda-1)^2(1-2\lambda^2 m_N(k,\theta)^2+3\lambda^2)s^2+(\lambda-1)^4s^4}{4} -\lambda^2 m_N(k,\theta)^2\sin^2(\alpha_{1,2})\right)^2\Bigg],
\end{multline}
where we have defined $s:=2n_\infty-1=(\gamma_+-\gamma_-)/(\gamma_++\gamma_-)$. Consequently, it seems challenging to obtain a closed form for generic $n$.

Let us now discuss the long-time term $\ln(\det \mathcal{Z'}_{n,\infty}^{(\boldsymbol{\alpha})})$. By using the eigenbasis of 
$\gcal'_\infty(k,t)$ one obtains 
\begin{multline}
\label{eq:det2_tinf}
        \det\left(\mathds{1}+\mathcal{W'}_\infty(k)\right)=\\
        =\det\left[\mathds{1}+\begin{pmatrix}
        p'&0&0&0\\0&q'&0&0\\0&0&p'^{-1}&0\\0&0&0&q'^{-1}
    \end{pmatrix}\exp\left(i\alpha_{1,2}\begin{pmatrix}
        \sigma_x+e\left(\sigma_z+i\sigma_y\right)&0_{2\times2}\\0_{2\times2}&\sigma_x+e\left(\sigma_z+i\sigma_y\right)
    \end{pmatrix}\right)...\right.\\
    \left....\begin{pmatrix}
        p'&0&0&0\\0&q'&0&0\\0&0&p'^{-1}&0\\0&0&0&q'^{-1}
    \end{pmatrix}\exp\left(i\alpha_{n,1}\begin{pmatrix}
        \sigma_x+e\left(\sigma_z+i\sigma_y\right)&0_{2\times2}\\0_{2\times2}&\sigma_x+e\left(\sigma_z+i\sigma_y\right)
    \end{pmatrix}\right)\right]
\end{multline}
where $\sigma_{x,y,z}$ are the Pauli matrices, and we defined $p':=(n'_+(k,\theta))/(1-n'_+(k,\theta))$, 
$q':=(n'_-(k,\theta))/(1-n'_-(k,\theta)/$, with
\begin{equation}
     n'_\pm(k,\theta,t):=\frac{1}{2}\left(1+\lambda(t)g_{12}(k,\theta)\pm \sqrt{\lambda(t)^2 f_{11}(k,\theta)^2+(1-\lambda(t))^2s^2}\right), 
\end{equation}
and
\begin{equation}
    \label{eq:e}
     e:=s(1-\lambda(t))/\sqrt{\lambda(t)^2 f_{11}(k,\theta)^2+(1-\lambda(t))^2s^2}.
\end{equation}
To obtain the determinant in~\eqref{eq:det2_tinf}, one can, again, exploit~\eqref{eq:det_to_tr_tinf}. Now, has to evaluate the moments
\begin{equation}
    \label{eq:n_moments'_tinf}
    M_\mathbf{h}^{(\infty)}:=\mathrm{Tr}\left[\begin{pmatrix}
        p&0\\0&q
    \end{pmatrix}\left(\sigma_x+e\left(\sigma_z+i\sigma_y\right)\right)^{h_1}...\begin{pmatrix}
        p&0\\0&q
    \end{pmatrix}\left(\sigma_x+e\left(\sigma_z+i\sigma_y\right)\right)^{h_L}\right]. 
\end{equation}
Unfortunately,  we are not able to  evaluate~\eqref{eq:n_moments'_tinf} in general, unlike the case 
with no dissipation (cf.~\eqref{eq:n_moments_tinf_intermediate}). Nevertheless, we can calculate $\ln( \det \mathcal{Z'}_{n,\infty}^{(\boldsymbol{\alpha})})$ for small values of $n$, by employing~\eqref{eq:n_moments_tinf_intermediate}. 
For example, for $n=2$ we have
\begin{equation}
  \label{eq:z2_generic_tinf_app}
  \ln\left(\det \mathcal{Z'}_{2,\infty}^{(\boldsymbol{\alpha})}\right)=2 \ln \left(h_2(n'_+(k,\theta,t))h_2(n'_-(k,\theta,t))-\lambda^2(t) f_{11}(k,t)^2\tilde{h}'_2(\boldsymbol{\alpha},k,\theta,t) \right),
\end{equation}
where $\tilde{h}'_n(\boldsymbol{\alpha},k,\theta,t)$ is obtained by substituting $f_{11}(k,\theta)\rightarrow \lambda(t)f_{11}(k,\theta)$ and $n_\pm(k,\theta) \rightarrow n'_\pm (k,\theta,t)$ in the definition~\eqref{eq:angles_tinf} of $\tilde{h}_n(\boldsymbol{\alpha},k,\theta)$. 

By using~\eqref{eq:z2_generic_tinf_app} and~\eqref{eq:z2_generic_t0_app} in~\eqref{eq:chmom_general_diss} we obtain the leading 
order of $\ln Z_{2}(\boldsymbol{\alpha},t)$ in the weakly-dissipative hydrodynamic limit. 

\section{Proof of Eq.~\eqref{eq:z2_XY}}\label{app:XY_proof}

Here we derive~\eqref{eq:z2_XY}. We consider the $XX$ chain (see section~\ref{sec:models}). 
Th pre-quench initial state is the ground state of the  $XY$ chain. As explained in section~\ref{sec:models}, 
only entangled pairs of quasiparticles are produced after the quench. Eq.~\eqref{eq:z2_XY} is 
the contribution to the charged moments $\ln(Z_{n}^{(\boldsymbol{\alpha})})$ due to the configuration in 
which both the quasiparticles forming the pair are in  subsystem $A$. 
To prove~\eqref{eq:z2_XY} one has to compute 
\begin{equation}
    \label{eq:dets}
    \det\left(\frac{\mathds{1}-G_{\{1,2\}}(k)}{2}\right)^n\det\left(\mathds{1}+\mathcal{W}_{\{1,2\}}(k)\right),
\end{equation}
where $\mathcal{W}_{\{1,2\}}(k):=\prod_{j=1}^{n}(\mathds{1}+G_{\{1,2\}}(k)))(\mathds{1}-G_{\{1,2\}}(k))^{-1}e^{i(\alpha_j-\alpha_{j+1})n_\mathcal{A}}$, with $G_{\{1,2\}}(k)$ defined in~\eqref{eq:G-supp} and $n_\mathcal{A}=\text{diag}(-1,-1,1,1)$. The subscript $\{1,2\}$ means that 
one has to restrict the correlator to the species of quasiparticles that are in $A$. Notice that since there are only two species of 
quasiparticles one has that $G_{\{1,2\}}(k)=G(k)$, i.e., the restricted correlator is the full 
correlator. Now, we can adapt the strategy used  in Appendix~\ref{app:proof_nodiss}, obtaining 
the equivalent of Eq.~\eqref{eq:cto1} as 
\begin{equation}
    \label{eq:XY_cto1}
    \det\left(\frac{\mathds{1}-G(k)}{2}\right)^n\det\left(\mathds{1}+\mathcal{W}_{\{1,2\}}(k)\right)=\frac{\mathrm{Tr}\left[\overline{\mathcal{W}_{\{1,2\}}(k)}\right]^2-\mathrm{Tr}\left[\overline{\mathcal{W}_{\{1,2\}}(k)}^2\right]}{2\cdot 4^n},
\end{equation}
where $\overline{\mathcal{W}_{\{1,2\}}(k)}:=\prod_{j=1}^{n}(\mathds{1}+G(k))e^{i(\alpha_j-\alpha_{j+1})n_\mathcal{A}}$. 
Again, we proceed by Taylor-expanding the exponentials $e^{i(\alpha_j-\alpha_{j+1})n_\mathcal{A}}$. Thus, one defines the moments 
$M_{\mathbf{h}}^{\{1,2\}}$ as 
\begin{equation}
    \label{eq:XY_n_moments}
    M_\mathbf{h}^{\{1,2\}}:=\mathrm{Tr}\left[(\mathds{1}+G(k))n_\mathcal{A}^{h_1}...(\mathds{1}+G(k))n_\mathcal{A}^{h_L}\right]. 
\end{equation}
Eq.~\eqref{eq:XY_n_moments} can be rewritten as 
\begin{equation}
    \label{eq:XY_n_moments_2}
    M_\mathbf{h}^{\{1,2\}}=2^{\#\text{ even }h_j}\mathrm{Tr}\left[(n_\mathcal{A}(\mathds{1}+G(k)))^{\#\text{ odd }h_j}\right].
\end{equation}
To perform the trace in~\eqref{eq:XY_n_moments_2} we diagonalize $n_\mathcal{A}(\mathds{1}+G(k))$.  The only  
doubly-degenerate eigenvalue is $2\cos \Delta_k$. Thus, we obtain 
\begin{equation}
    \label{eq:XY_n_moments_3}
    M_\mathbf{h}^{\{1,2\}}=2^{L+1} \left(\cos \Delta_k\right)^{\sum_{j=1}^L \frac{1-(-1)^{h_j}}{2}}.
\end{equation}
We then sum over $h_j$, reconstructing the exponentials of $n_\mathcal{A}$. Each  term yields a factor $\cos\alpha_{j,j+1}+ i \cos \Delta_k \sin\alpha_{j,j+1}=f_k(\alpha_{j,j+1})$.  We obtain 
\begin{align}
    \label{eq:XY_tracesW12}
        & \mathrm{Tr}\left[\overline{\mathcal{W}_{\{1,2\}}(k)}\right]=2^{n+1}\prod_{j=1}^n f_k(\alpha_{j,j+1}),\\
        \label{eq:XY_tracesW12-2}
        & \mathrm{Tr}\left[\overline{\mathcal{W}_{\{1,2\}}(k)}^2\right]=2^{2n+1}\prod_{j=1}^n f_k(\alpha_{j,j+1})^2.
\end{align}
Finally, by substituting~\eqref{eq:XY_tracesW12} and~\eqref{eq:XY_tracesW12-2} into~\eqref{eq:XY_cto1}, 
and taking the (halved) logarithm, we obtain the desired Eq.~\eqref{eq:z2_XY}. 

\end{appendices}

\end{document}